\documentclass[10pt]{article}
\usepackage{bbm}
\usepackage[final]{graphics}
\usepackage{amsmath,amscd}
\usepackage{amsfonts,amsbsy}
\usepackage{amssymb}
\usepackage[scaled=0.92]{helvet}
\usepackage{color}

\def\empile#1\over#2{\mathrel{\mathop{\kern 0pt#1}\limits_{#2}}}
\def\bs{\boldsymbol}
\def\wt#1{\widetilde{#1}}

\newcommand{\slv}{\raise.15ex\hbox{$/$}\kern-.53em\hbox{$v$}}
\newcommand{\slF}{\raise.15ex\hbox{$/$}\kern-.53em\hbox{$F$}}
\newcommand{\slL}{\raise.15ex\hbox{$/$}\kern-.53em\hbox{$L$}}
\newcommand{\slP}{\raise.15ex\hbox{$/$}\kern-.53em\hbox{$P$}}
\newcommand{\slp}{\raise.15ex\hbox{$/$}\kern-.53em\hbox{$p$}}
\newcommand{\slq}{\raise.15ex\hbox{$/$}\kern-.53em\hbox{$q$}}
\newcommand{\slR}{\raise.15ex\hbox{$/$}\kern-.53em\hbox{$R$}}
\newcommand{\slQ}{\raise.15ex\hbox{$/$}\kern-.53em\hbox{$Q$}}
\newcommand{\slK}{\raise.15ex\hbox{$/$}\kern-.53em\hbox{$K$}}
\newcommand{\slk}{\raise.15ex\hbox{$/$}\kern-.53em\hbox{$k$}}
\newcommand{\slD}{\raise.15ex\hbox{$/$}\kern-.73em\hbox{$D$}}
\newcommand{\slC}{\raise.15ex\hbox{$/$}\kern-.53em\hbox{$C$}}
\newcommand{\slA}{\raise.15ex\hbox{$/$}\kern-.53em\hbox{$A$}}
\newcommand{\slSigma}{\raise.15ex\hbox{$/$}\kern-.53em\hbox{$\Sigma$}}
\newcommand{\slpartial}{\raise.15ex\hbox{$/$}\kern-.53em\hbox{$\partial$}}
\newcommand{\slcalP}{\raise.15ex\hbox{$/$}\kern-.63em\hbox{$\cal P$}}

\def\p{{\boldsymbol p}}
\def\q{{\boldsymbol q}}

\def\k{{\boldsymbol k}}

\def\x{{\boldsymbol x}}
\def\y{{\boldsymbol y}}

\def\z{{\boldsymbol z}}

\def\u{{\boldsymbol u}}

\def\rmd{{\rm d}}

\catcode`\@=11

\newcount\@tempcntc
\def\@citex[#1]#2{\if@filesw\immediate\write\@auxout{\string\citation{#2}}\fi
  \@tempcnta\z@\@tempcntb\m@ne\def\@citea{}\@cite{%
        \@for\@citeb:=#2\do%
    {\@ifundefined{b@\@citeb}%
        {\@citeo\@tempcntb\m@ne\@citea%
                \def\@citea{,\penalty\@m\ }{\bf ?}\@warning%
                {Citation `\@citeb' on page \thepage \space undefined}}%
        {\setbox\z@\hbox{\global\@tempcntc0\csname b@\@citeb\endcsname\relax}
     \ifnum\@tempcntc=\z@ \@citeo\@tempcntb\m@ne%
       \@citea\def\@citea{,\penalty\@m}%
       \hbox{\csname b@\@citeb\endcsname}%
     \else%
      \advance\@tempcntb\@ne%
      \ifnum\@tempcntb=\@tempcntc%
      \else\advance\@tempcntb\m@ne\@citeo%
      \@tempcnta\@tempcntc\@tempcntb\@tempcntc\fi\fi}}\@citeo}{#1}}%

\def\@citeo{\ifnum\@tempcnta>\@tempcntb\else\@citea
  \def\@citea{,\penalty\@m}%
  \ifnum\@tempcnta=\@tempcntb\the\@tempcnta\else
   {\advance\@tempcnta\@ne\ifnum\@tempcnta=\@tempcntb \else
\def\@citea{--}\fi
    \advance\@tempcnta\m@ne\the\@tempcnta\@citea\the\@tempcntb}\fi\fi}

\catcode`\@=12

\begin{document}

\title{\bf Tree-level correlations in\\the strong field regime} \author{Fran\c cois Gelis} \maketitle

\begin{center}
  Institut de physique th\'eorique\\
  CEA, CNRS, Universit\'e Paris-Saclay\\
  F-91191 Gif-sur-Yvette, France
\end{center}
\vglue 10mm

\begin{abstract}
  We consider the correlation function of an arbitrary number of local
  observables in quantum field theory. We show that, at tree level in
  the strong field regime, these correlations arise solely from
  fluctuations in the initial state. We obtain the general expression
  of these correlation functions in terms of the classical solution of
  the field equation of motion and its derivatives with respect to its
  initial conditions, that can be arranged graphically as the sum of
  labeled trees where the nodes are the individual observables, and
  the links are pairs of derivatives acting on them. For $3$-point
  (and higher) correlation functions, there are additional tree-level
  terms beyond from the strong field approximation, generated
  throughout the evolution of the system.
\end{abstract}

\section{Introduction}
A common question in many areas is the evaluation of correlations
between several measurements, given the microscopic dynamics of the
system. This problem comes up for instance in the calculation of
cosmological perturbations \cite{Mukhanov:1990me,Weinberg:2005vy}, or
in nuclear physics, e.g. in heavy ion collisions
\cite{Gelis:2010nm,Gelis:2012ri}. These situations have in common that
they are described by some underlying quantum field theory, and that
the system starts from some supposedly known initial state. Then, it
evolves under the effect of the self-interactions of the fields, and
possibly the couplings to some external sources. Thereafter, we define
an observable as some local operator ${\cal O}(\phi(x))$ constructed
from the fields of the theory. Our goal in this paper is to study (at
leading order in the couplings) the correlations between measurements
of this observable at several space-time points $x_1,x_2,\cdots,x_n$,
\begin{equation}
{\cal C}_{\{1\cdots n\}}\equiv\big<0{}_{\rm in}\big|{\cal O}(\phi(x_1))\cdots {\cal O}(\phi(x_n))\big|0{}_{\rm in}\big>_{\rm c}\; . 
\label{eq:corr-def}
\end{equation}
In this definition, we have assumed that the system is initially in
the perturbative vacuum state (see the appendices \ref{app:coherent}
and \ref{app:mixed} for a discussion of the changes with other types
of initial states).  The subscript ``c'' indicates the connected part
of this expectation value, i.e. the terms that contain the actual
correlations. An important restriction in our discussion will be to
consider only points $x_i$ that all have space-like separations,
i.e. $(x_i-x_j)^2<0$. Physically, this means that the measurement
performed at one of the points cannot influence the outcome of the
measurement at another point. Therefore, the correlations between the
measurements are entirely due to the past evolution of the system.

The main difference compared to the more familiar problem of computing
cross-sections is that the final state is not prescribed. Instead, one
wishes to calculate the expectation value of some operators, by
summing over {\sl all the possible final states} given a prescribed
initial state\footnote{This could be generalized somehow, by having a
  mixed initial state described by some density matrix, rather than a
  pure state. See the appendix \ref{app:mixed}.}. One could in
principle perform such a calculation by using the usual techniques for
calculating transition amplitudes, by first writing
\begin{equation}
  \big<{\rm in}\big|{\cal O}\big|{\rm in}\big>
  =\sum_{{\rm f},{\rm f}'}\big<{\rm in}\big|{\rm f}\big>
  \big<{\rm f}\big|{\cal O}\big|{\rm f}'\big>\big<{\rm f}'\big|{\rm in}\big>\; .
\end{equation}
(The double sum over final states could be reduced to a single sum if
we use eigenstates of the observable ${\cal O}$.) All the steps
involved in this pedestrian approach can be encapsulated in a set of
diagrammatic rules known as the Schwinger-Keldysh formalism, or in-in
formalism
\cite{Schwinger1961,Bakshi:1962dv,Bakshi:1963bn,Keldysh:1964ud} (see
also \cite{Chou:1984es,Jordan:1986ug}), that roughly consists in two
copies of the usual Feynman rules (one that gives the factor
$\big<{\rm f}'\big|{\rm in}\big>$, and one --complex conjugated-- that
gives the factor $\big<{\rm in}\big|{\rm f}\big>$), plus some
additional rules that give the final state sum. The approach we follow
in this paper bears heavily on earlier works in the field of heavy ion
collisions \cite{Gelis:2006yv,Gelis:2006cr,Gelis:2008rw,Gelis:2008ad}
and in cosmology \cite{Weinberg:2005vy,Weinberg:2008mc}.

Although the approach used in this paper and the final result are
quite general, the intermediate derivations are a bit cumbersome. In
order to keep the notations as light as possible, we consider a theory
with a single real scalar field $\phi$, whose Lagrangian density is given
by
\begin{equation}
{\cal L}\equiv \frac{1}{2}(\partial_\mu \phi)(\partial^\mu \phi)-\frac{1}{2}m^2\,\phi^2 \underbrace{-V(\phi) + J\phi}_{{\cal L}_{\rm int}(\phi)}\; ,
\label{eq:lag}
\end{equation}
where $V(\phi)$ contains the self-interactions of the field and $J(x)$
is an external source. The quantum field theory described by the
Lagrangian density (\ref{eq:lag}) has a well known perturbative
expansion. However, we are interested in this paper in the {\sl strong
  field regime}, where this perturbative expansion is
insufficient. This corresponds to fields for which the interaction
term $V(\phi)$ is as large as the kinetic term
$(\partial_\mu\phi)(\partial^\mu\phi)$,
\begin{equation}
(\partial_\mu \phi)(\partial^\mu \phi)\sim V(\phi)\; ,
\end{equation}
implying that the interactions cannot be treated as a
perturbation. For instance, for a theory with a quartic interaction
term $V(\phi)\sim g^2\phi^4$, this occurs for fields
\begin{equation}
\phi\sim \frac{Q}{g}\; ,
\end{equation}
where $Q$ is some typical momentum scale in the problem (the field
$\phi$ has the dimension of a momentum in four space-time
dimensions). For fields driven by an external source, the term $J\phi$
must also have the same order of magnitude as the kinetic and
self-interaction terms, and thus the order of magnitude of the source
that may create this large field\footnote{Alternatively, instead of
  creating large fields by the coupling to an external source, one may
  consider a system initialized in a coherent state where the field is
  initially large.} is
\begin{equation}
J\sim \frac{Q^3}{g}\; .
\label{eq:largeJ}
\end{equation}
In order to see where the usual perturbative expansion fails, let us
recall the power counting in the case of a theory with a $g^2\phi^4$
interaction term. The order of magnitude of a connected graph
${\cal G}$ with $n_{_E}$ external legs, $n_{_L}$ loops and $n_{_J}$
external sources is given by
\begin{equation}
{\cal G}\sim g^{-2} g^{-n_{_E}}g^{-2n_{_L}} \big(gJ\big)^{n_{_J}}\; .
\end{equation}
Eq.~(\ref{eq:largeJ}) implies that $gJ$ cannot be treated as a small
expansion parameter. Any quantity must therefore be evaluated
non-perturbatively in the number of sources $n_{_J}$. In contrast, the
hierarchy of the contributions based on the number of loops in the
Feynman diagrams survives. Our aim is to calculate the correlation
function (\ref{eq:corr-def}) to lowest order in the number of loops
(i.e. at tree level) but to all order in $n_{_J}$ (or equivalently to
all orders in the field amplitude).

With large fields, the correlation function (\ref{eq:corr-def}) is
suppressed compared to the uncorrelated part, because each propagator
connecting a pair of observables costs a factor $g^2$ (the endpoints
of such a link replace two fields of order $g^{-1}$). In order to
fully connect the $n$ observables, $n-1$ propagators are necessary,
suppressing the correlation by a factor $g^{2(n-1)}$ with respect to
the uncorrelated part. Thus, in a certain sense, the correlation
function may be viewed as a higher order correction. At leading order,
$1$-point functions (i.e. the expectation value of a local observable)
are obtained from a classical solution of the field equation of
motion, obeying a retarded boundary condition that depends on the
initial state ($\phi=\dot\phi=0$ when the initial state is the
perturbative vacuum) \cite{Gelis:2006yv}. Their next-to-leading order
(NLO) corrections are also known \cite{Gelis:2008rw}, and can be
expressed in terms of functional derivatives of this classical field
with respect to its initial condition. A similar result was proven in
\cite{Gelis:2008ad} for the NLO correction to a $2$-point function,
that contains its tree-level correlated part.

In this paper, we assess to what extent a similar representation
exists for the correlated part of higher $n$-point functions at
tree-level.  We prove that, within the {\sl strong field
  approximation} (defined in the section \ref{sec:strong}), the
tree-level correlation between any number of arbitrary local
observables takes the following graphical form: \setbox1\hbox to
37.5mm{\resizebox*{37.5mm}{!}{\includegraphics{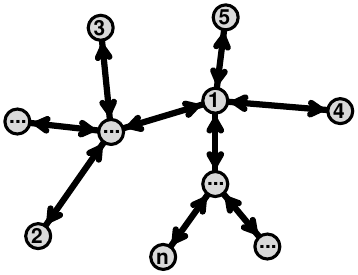}}}
\begin{equation}
 {\cal C}_{\{1\cdots n\}} =\sum_{{\mbox{\scriptsize trees with }n}\atop{\mbox{\scriptsize labeled  nodes}}}\raise -17mm\box1\; ,
  \label{eq:Cn-trees-0}
\end{equation}
where the nodes are the observables evaluated for a classical solution
of the field equation of motion, and the links are pairs of
derivatives with respect to the initial condition of the classical
field.

The paper is organized as follows. In the section \ref{sec:gen}, we
couple the observable ${\cal O}(\phi(x))$ to a fictitious source
$z(\x)$ in order to define a generating functional whose derivatives
give the expectation for the combined measurement of
${\cal O}(\phi(x))$ at several points. We explain how to obtain this
generating functional diagrammatically by extending the
Schwinger-Keldysh formalism with an extra vertex that corresponds to
the observable, and then we rephrase this diagrammatic expansion in
the retarded-advanced basis, that will be useful to discuss the strong
field regime. In the section \ref{sec:1der}, we study the first
derivative of the generating functional. At leading order, it can be
expressed in terms of a pair of fields that obey classical equations
of motion and satisfy non-trivial $z$-dependent boundary conditions.
In the following section \ref{sec:expansion}, we set up an expansion
of these fields order by order in powers of the function $z(\x)$, and
we determine explicitly the first two orders, that give the $1$-point
and $2$-point correlation functions. We also show how to rewrite the
$2$-point function in terms of derivatives of a classical field with
respect to its initial condition.  In the section \ref{sec:strong}, we
consider the strong field approximation, and we determine the solution
to all orders in $z$ in this approximation. We show that the tree
level $n$-point correlation function in this approximation can be
represented as the sum of all trees with $n$ labeled nodes (the $n$
instances of the observable), and pairwise links that are differential
operators with respect to the initial condition of the classical
field. The section \ref{sec:beyond} illustrates on the example of the
$3$-point function the type of contributions that arise at tree level
beyond the strong field approximation. Summary and conclusions are in
the section \ref{sec:concl}. In the appendix \ref{app:modes}, we
derive some technical points used in the main part of the paper, while
in the appendices \ref{app:coherent} and \ref{app:mixed} we discuss
other types of initial states.

\section{Generating functional for local measurements}
\label{sec:gen}
\subsection{Definition}
For simplicity, we will take the points $x_i$ where the measurements
are performed to lie on the same surface of constant time $x^0=t_f$,
but the final results would be valid for any locally space-like
surface (in order to ensure that there is no causal relation between
the points $x_i$).

One can encapsulate all the correlation functions
(\ref{eq:corr-def}) into a generating functional defined as
follows\footnote{This is the definition for the case where the initial
  state is the vacuum. The changes to the formalism when the initial
  state is a coherent state are discussed in the appendix
  \ref{app:coherent}, and the case of a Gaussian mixed state is
  discussed in the appendix \ref{app:mixed}.}:
\begin{equation}
{\cal F}[z(\x)]
\equiv
\big<0{}_{\rm in}\big|
\exp\int_{t_f} \rmd^3\x\;z(\x)\,{\cal O}(\phi(x))\big|0{}_{\rm in}\big>\; ,
\label{eq:F-def}
\end{equation}
where the argument of the field in ${\cal O}$ is $x\equiv (t_f,\x)$.
From this generating functional, the correlation functions are
obtained by differentiating with respect to $z(\x_1),\cdots,z(\x_n)$ and
by setting $z\equiv 0$ afterwards. In order to remove the uncorrelated
part of the $n$-point function, we should differentiate the logarithm
of ${\cal F}$, i.e.
\begin{equation}
{\cal C}_{\{1\cdots n\}}
=
\left.\frac{\delta^n \ln{\cal F}}{\delta z(\x_1)\cdots \delta z(\x_n)}\right|_{z\equiv 0}
\end{equation}
Note that since the final surface is space-like and the operators
${\cal O}(\phi(x))$ are local, there is no need for a specific
ordering of the exponential. The observable ${\cal O}(\phi(x))$ is
made of the field in the Heisenberg picture, $\phi(x)$, that can be
related to the field $\phi_{\rm in}(x)$ of the interaction picture as
follows:
\begin{equation}
\phi(x)=U(-\infty,x^0)\,\phi_{\rm in}(x)\,U(x^0,-\infty)\; ,
\end{equation}
where $U(t_1,t_2)$ is an evolution operator given in terms of the interactions
by the following formula
\begin{equation}
U(t_2,t_1)={\rm T}\,\exp i\int_{t_1}^{t_2} \rmd x^0 \rmd^3\x\;{\cal L}_{\rm int}(\phi_{\rm  in}(x))\; .
\end{equation}
We can therefore rewrite the generating functional solely in terms of
the interaction picture field $\phi_{\rm in}$,
\begin{eqnarray}
{\cal F}[z(\x)]
&=&
\big<0{}_{\rm in}\big|
{\rm P}\,\exp \int \rmd^3\x\; \Big\{
i\int \rmd x^0\;{\cal L}_{\rm int}(\phi_{{\rm in} +}(x))-{\cal L}_{\rm int}(\phi_{{\rm in} -}(x))
\nonumber\\
&&\qquad\qquad\qquad\qquad\qquad
+z(\x)\,{\cal O}(\phi_{\rm in}(t_f,\x))
\Big\}
\big|0{}_{\rm in}\big>\; ,
\label{eq:Fin}
\end{eqnarray}
where ${\rm P}$ denotes a {\sl path ordering} on the following time
contour: \setbox1\hbox to
8cm{\resizebox*{8cm}{!}{\includegraphics{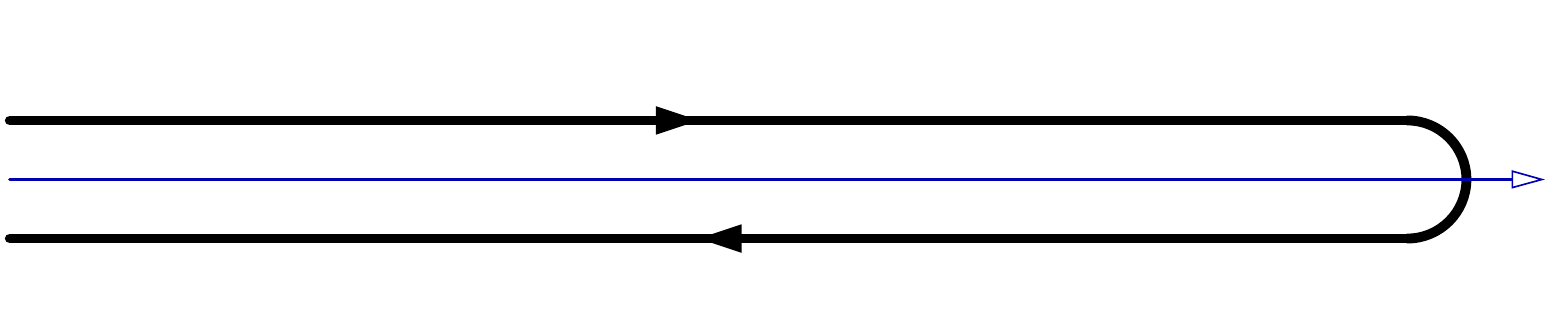}}}
\begin{equation*}
  \resizebox*{7cm}{!}{\begin{picture}(0,0)%
\includegraphics{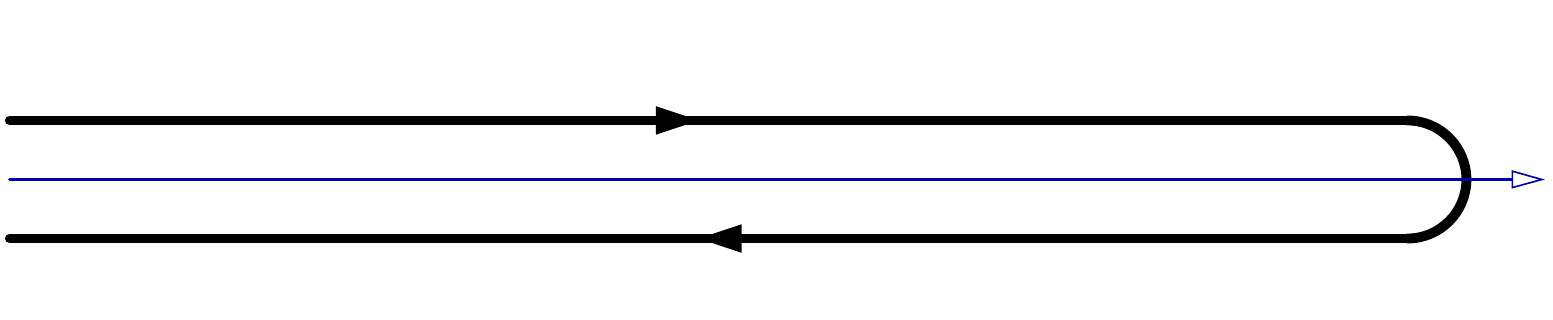}%
\end{picture}%
\setlength{\unitlength}{4144sp}%
\begingroup\makeatletter\ifx\SetFigFontNFSS\undefined%
\gdef\SetFigFontNFSS#1#2#3#4#5{%
  \reset@font\fontsize{#1}{#2pt}%
  \fontfamily{#3}\fontseries{#4}\fontshape{#5}%
  \selectfont}%
\fi\endgroup%
\begin{picture}(7086,1447)(497,-3311)
\put(991,-2131){\makebox(0,0)[lb]{\smash{{\SetFigFontNFSS{20.74}{24.0}{\familydefault}{\mddefault}{\updefault}{\color[rgb]{0,0,0}${\cal C}$}%
}}}}
\put(7471,-2491){\makebox(0,0)[lb]{\smash{{\SetFigFontNFSS{20.74}{24.0}{\familydefault}{\mddefault}{\updefault}{\color[rgb]{0,0,0}$x^0$}%
}}}}
\put(4411,-3211){\makebox(0,0)[lb]{\smash{{\SetFigFontNFSS{20.74}{24.0}{\familydefault}{\mddefault}{\updefault}{\color[rgb]{0,0,0}$-$}%
}}}}
\put(4411,-2221){\makebox(0,0)[lb]{\smash{{\SetFigFontNFSS{20.74}{24.0}{\familydefault}{\mddefault}{\updefault}{\color[rgb]{0,0,0}$+$}%
}}}}
\end{picture}%
}
\end{equation*}
The operators are ordered from left to right starting from the end of
the contour. We denote by $\phi_{{\rm in}+}$ the field that lives on
the upper branch and by $\phi_{{\rm in}-}$ the field on the lower
branch (the minus sign in front of the term
${\cal L}_{\rm int}(\phi_{{\rm in} -}(x))$ comes from the fact that
the lower branch is oriented from $+\infty$ to $-\infty$).  The
operator ${\cal O}(\phi_{\rm in}(x))$ lives at the final time of this
contour, and could either be viewed as made of fields of type $+$ or
of type $-$ (the two choices lead to the same results). Note that in
eq.~(\ref{eq:Fin}), the spacetime integration is a priori extended to
the domain located below the final surface (the observable ${\cal O}$
therefore lives on the upper time boundary of the integration
domain). However, this restriction is not really necessary: by
causality, the contribution of the domain located above $t_f$
would cancel anyhow.

\subsection{Expression in the Schwinger-Keldysh formalism}
\begin{figure}[htbp]
\begin{center}
\resizebox*{6cm}{!}{\includegraphics{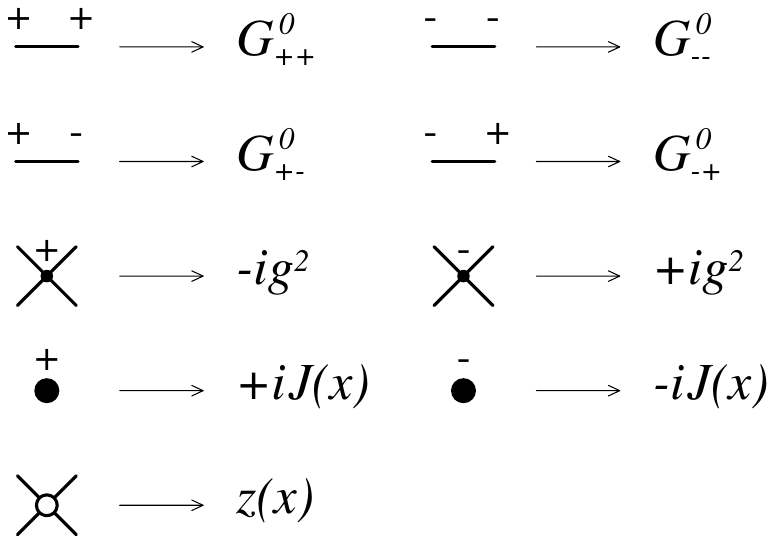}}
\end{center}
\caption{\label{fig:rules}Diagrammatic rules for the extended
  Schwinger-Keldysh formalism that gives the generating
  functional. The Feynman rules shown here for the self-interactions
  correspond to a $g^2\phi^4/4!$ interaction term. In this
  illustration, we have assumed that the observable is quartic in the
  field when drawing the corresponding vertex (proportional to
  $z(\x)$).}
\end{figure}

Since the initial state is the vacuum, the generating functional
defined in eq.~(\ref{eq:Fin}) can be represented diagrammatically as
the sum of all the vacuum-to-vacuum graphs (i.e. graphs without
external legs) in the Schwinger-Keldysh formalism (also known as the
{\sl in-in} formalism), extended by an extra vertex that corresponds
to the insertions of the observable ${\cal O}$. Let us recall here
that the Schwinger-Keldysh diagrammatic rules consist in having two
types of interaction vertices ($+$ and $-$ depending on which branch
of the contour the vertex lies on, the $-$ vertex being the opposite
of the $+$ one) and four types of bare propagators
($G_{++}^0, G_{--}^0,G_{+-}^0$ and $G_{-+}^0$) depending on the
location of the endpoints on the contour. The additional vertex exists
only on the final surface, at the time $t_f$. It is accompanied by a
factor $z(\x)$, and has as many legs as there are fields in ${\cal
  O}$. There is only one kind of this vertex (we can decide to call it
$+$ or $-$ without affecting anything). We recapitulate these Feynman
rules in the figure \ref{fig:rules}.
In the case of the vacuum initial state, the  propagators have the
following explicit expressions:
\begin{eqnarray}
  &&
  G_{-+}^0(x,y)=\int_\k e^{-ik\cdot(x-y)}
  \;,\quad
  G_{+-}^0(x,y)=\int_\k e^{ik\cdot(x-y)}
     \; ,\nonumber\\
  && G_{++}^0(x,y)=\theta(x^0-y^0)\,G_{-+}^0(x,y)+\theta(y^0-x^0)\,G_{+-}^0(x,y)\; ,
     \nonumber\\
  && G_{--}^0(x,y)=\theta(x^0-y^0)\,G_{+-}^0(x,y)+\theta(y^0-x^0)\,G_{-+}^0(x,y)\; ,
     \label{eq:props-vac}
\end{eqnarray}
where we have used the following compact notation
\begin{equation}
\int_\k\cdots\equiv \int \frac{\rmd^3\k}{(2\pi)^3 2E_\k}\cdots\qquad (E_\k\equiv\sqrt{\k^2+m^2})\; .
\end{equation}

Note that when we set $z\equiv 0$, these
diagrammatic rules fall back to the pure Schwinger-Keldysh formalism,
for which it is known that all the connected vacuum-to-vacuum graphs
are zero. This implies that
\begin{equation}
{\cal F}[z\equiv 0]=1\; ,
\end{equation}
in accordance with the fact that this should be
$\big<0{}_{\rm in}\big|0{}_{\rm in}\big>=1$.

\subsection{Retarded-advanced representation}
In order to clarify what approximations may be done in the large field
regime, it is useful to use a different basis of fields by introducing \cite{Aurenche:1993vt,vanEijck:1994rw}
\begin{equation}
  \phi_2\equiv \tfrac{1}{2}\,\big(\phi_++\phi_-\big)\; ,\qquad
  \phi_1\equiv \phi_+-\phi_-\; .
\end{equation}
The half-sum $\phi_2$ in a sense captures the classical content (plus
some quantum corrections), while the difference $\phi_1$ is purely
quantum (because it represents the different histories of the fields
in the amplitude and in the complex conjugated amplitude). To see
how the Feynman rules are modified in terms of these new fields, let us
start from
\begin{equation}
  \phi_\alpha=\sum_{\epsilon=\pm}\Omega_{\alpha\epsilon}\,\phi_\epsilon\qquad(\alpha=1,2)\; ,
  \label{eq:SK-rotation}
\end{equation}
where the matrix $\Omega$ reads:
\begin{equation}
\Omega_{\alpha\epsilon}
\equiv
\begin{pmatrix}
1 & -1 \\
1/2 & 1/2 \\
\end{pmatrix}\; .
\end{equation}
In terms of this matrix, the new propagators are obtained as follows
\begin{eqnarray}
{G}_{\alpha\beta}^0\equiv 
\sum_{\epsilon,\epsilon^\prime=\pm}
\Omega_{\alpha\epsilon}\Omega_{\beta\epsilon^\prime}
{G}_{\epsilon\epsilon^\prime}^0\; .
\label{eq:SK-rotation-prop}
\end{eqnarray}
Explicitly, these propagators read
\begin{eqnarray}
  G_{21}^0&=& G_{++}^0-G_{+-}^0\; ,\nonumber\\
  G_{12}^0&=& G_{++}^0-G_{-+}^0\; ,\nonumber\\
  G_{22}^0&=& \tfrac{1}{2}\,\left[G_{+-}^0+G_{-+}^0\right]\; ,\nonumber\\
  G_{11}^0&=& 0\; .
\end{eqnarray}
Note that $G_{21}^0$ is the bare retarded propagator, while $G_{12}^0$
is the bare advanced propagator. The vertices in the new formalism
(here written for a quartic interaction) are given by
\begin{equation}
\Gamma_{\alpha\beta\gamma\delta}\equiv -ig^2\left[
  \Omega^{-1}_{+\alpha}\Omega^{-1}_{+\beta}\Omega^{-1}_{+\gamma}\Omega^{-1}_{+\delta}
  -
  \Omega^{-1}_{-\alpha}\Omega^{-1}_{-\beta}\Omega^{-1}_{-\gamma}\Omega^{-1}_{-\delta}
\right]\; ,
\end{equation}
where
\begin{equation}
\Omega^{-1}_{\epsilon\alpha}=
\begin{pmatrix}
1/2 & 1 \\
-1/2 & 1 \\
\end{pmatrix}\qquad\qquad[\Omega_{\alpha\epsilon}\Omega_{\epsilon\beta}^{-1}=\delta_{\alpha\beta}]\; .
\end{equation}
More explicitly, we have~:
\begin{eqnarray}
&&\Gamma_{1111}=\Gamma_{1122}=\Gamma_{2222}=0\nonumber\\
&&\Gamma_{1222}=-ig^2\; ,\quad \Gamma_{1112}=-ig^2/4\; .
\end{eqnarray}
(The vertices not listed explicitly here are obtained by
permutations.)  Finally, the rules for an external source
in the retarded-advanced basis are~:
\begin{equation}
J_1=J\;,\quad J_2 = 0\; .
\end{equation}
Finally, note that the observable depends only on the field $\phi_2$,
i.e. ${\cal O}={\cal O}(\phi_2)$. Indeed, the fields $\phi_+$ and
$\phi_-$ represent the field in the amplitude and in the conjugated
amplitude. Their difference should vanish when a measurement is
performed.

\section{First derivative at tree level}
\label{sec:1der}
\subsection{First derivative of $\ln{\cal F}$}
Differentiating the generating functional with respect to $z(\x)$
amounts to exhibiting a vertex ${\cal O}$ at the point $\x$ at the
final time (as opposed to weighting this vertex by $z(\x)$ and
integrating over $\x$). Furthermore, by considering the logarithm of
the generating functional rather than ${\cal F}$ itself, we have only
diagrams that are connected to the point $\x$, as shown in this
representation: \setbox1\hbox to
25mm{\resizebox*{25mm}{!}{\includegraphics{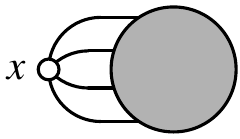}}}
\begin{equation}
\frac{\delta \ln{\cal F}}{\delta z(\x)}
=\;\raise -6.4mm\box1\;\; ,
\label{eq:1der}
\end{equation}
where the gray blob is a sum of graphs constructed with the Feynman
rules of the figure \ref{fig:rules}, or their analogue in the
retarded-advanced formulation. Therefore, these graphs still depend
implicitly on $z$. Note that this blob does not have to be connected.

\subsection{Tree level expression}
Without further specifying the content of the blob,
eq.~(\ref{eq:1der}) is valid to all orders, both in $z$ and in $g$. At
lowest order in $g$ (tree level), a considerable simplification happens
because the blob must be a product of disconnected subgraphs, one for
each line attached to the vertex ${\cal O}(\phi(x))$: \setbox1\hbox to
20mm{\resizebox*{20mm}{!}{\includegraphics{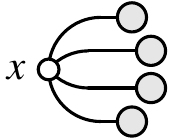}}}
\begin{equation}
\left.\frac{\delta \ln{\cal F}}{\delta z(\x)}\right|_{\rm tree}
=\;\raise -7.1mm\box1\;\; ,
\label{eq:1der-tree}
\end{equation}
where now each of the light colored blob is a {\sl connected tree}
$1$-point diagram. In the retarded-advanced formalism, there are two
of these $1$-point functions, that we will denote $\phi_1$ and
$\phi_2$. At tree level, they can be defined recursively by the
following pair of coupled integral equations:
\begin{eqnarray}
\phi_1(x)
&=&
i\int_\Omega \rmd^4y\;
    G_{12}^0(x,y)\,\frac{\partial{\bs L}_{\rm int}(\phi_1,\phi_2)}{\partial \phi_2(y)}
\nonumber\\
&&\qquad
+\int_{t_f} \rmd^3\y\;
G_{12}^0(x,y)\;z(\y)\;{\cal O}'(\phi_2(y))\; ,\nonumber\\
\phi_2(x)
&=&
i\int_\Omega \rmd^4y\;\Big\{
G_{21}^0(x,y)\,\frac{\partial{\bs L}_{\rm int}(\phi_1,\phi_2)}{\partial \phi_1(y)}
+
G_{22}^0(x,y)\,\frac{\partial{\bs L}_{\rm int}(\phi_1,\phi_2)}{\partial \phi_2(y)}
\Big\}\nonumber\\
&&\qquad
+\int_{t_f} \rmd^3\y\;
G_{22}^0(x,y)\;z(\y)\;{\cal O}'(\phi_2(y))\; .
\label{eq:int}
\end{eqnarray}
In these equations, ${\cal O}'$ is the derivative of the observable
with respect to the field, $\Omega$ is the space-time domain comprised
between the initial and final times, and we denote
\begin{equation}
  {\bs L}_{\rm int}(\phi_1,\phi_2)\equiv {\cal L}_{\rm int}(\phi_2+\tfrac{1}{2}\phi_1)-{\cal L}_{\rm int}(\phi_2-\tfrac{1}{2}\phi_1)\; .
  \label{eq:LL}
\end{equation}
For an interaction Lagrangian $-\tfrac{g^2}{4!}\phi^4+J\phi$, this
difference reads
\begin{equation}
  {\bs L}_{\rm int}(\phi_1,\phi_2)
  =
  -\frac{g^2}{6}\phi_2^3\phi_1 -\frac{g^2}{4!}\phi_1^3\phi_2 +J\phi_1\; .
  \label{eq:LL-phi4}
\end{equation}
  In terms of these fields,
we have
\begin{equation}
\left.\frac{\delta \ln{\cal F}}{\delta z(\x)}\right|_{\rm tree}
={\cal O}(\phi_2(x))\; ,
\label{eq:1der-tree-1}
\end{equation}
i.e. simply the observable ${\cal O}$ evaluated on the field
$\phi_2(x)$ (but this field depends on $z$ to all orders, via the
boundary terms in eqs.~(\ref{eq:int})).

\subsection{Classical equations of motion}
Using the fact that $G_{12}^0$ and $G_{21}^0$ are Green's functions of
$\square+m^2$, respectively obeying the following identities
\begin{equation}
(\square_x+m^2)\,G_{12}^0(x,y)=-i\delta(x-y)\;,\quad
(\square_x+m^2)\,G_{21}^0(x,y)=+i\delta(x-y)\; ,
\end{equation}
while $G_{22}^0$ vanishes when acted upon by this operator,
\begin{equation}
(\square_x+m^2)\,G_{22}^0(x,y)=0\; ,
\end{equation}
we see that $\phi_1$ and $\phi_2$ obey the following classical field
equations of motion:
\begin{eqnarray}
  &&(\square_x+m^2)\,\phi_1(x)=
\frac{\partial{\bs L}_{\rm int}(\phi_1,\phi_2)}{\partial \phi_2(x)}\;,\nonumber\\
&&(\square_x+m^2)\,\phi_2(x)=\frac{\partial{\bs L}_{\rm int}(\phi_1,\phi_2)}{\partial \phi_1(x)}\; .
\label{eq:eoms}
\end{eqnarray}
Note that here the point $x$ is located in the ``bulk'' $\Omega$; this is
why the observable does not enter in these equations of motion. In
fact, the observable enters only in the boundary conditions satisfied
by these fields on the hypersurface at $t_f$.  For later reference,
let us also rewrite these equations of motion in the specific
case of a scalar field theory with a $g^2\phi^4/4!$ interaction term
and an external source $J$:
\begin{eqnarray}
  &&\Big[\square_x+m^2+\tfrac{g^2}{2}\phi_2^2\Big]\,\phi_1+\frac{g^2}{4!}\,\phi_1^3=0\;,\nonumber\\
&&(\square_x+m^2)\,\phi_2+\frac{g^2}{6}\,\phi_2^3+\frac{g^2}{8}\,\phi_1^2\phi_2=J\; .
\label{eq:eoms-phi4}
\end{eqnarray}

\subsection{Boundary conditions}
The equations of motion (\ref{eq:eoms}) are easier to handle than the
integral equations (\ref{eq:int}), but they must be supplemented with
boundary conditions in order to define uniquely the solutions. The
standard procedure for deriving the boundary conditions is to consider
the combination $G_{12}^0(x,y)\,({\square}_y+m^2)\,\phi_1(y)$, and let
the operator $\square_y+m^2$ act alternatively on the right and on the
left,
\begin{eqnarray}
&&
   G_{12}^0(x,y)\,(\stackrel{\rightarrow}{\square}_y+m^2)\,\phi_1(y)=G_{12}^0(x,y)\,
   \frac{\partial{\bs L}_{\rm int}(\phi_1,\phi_2)}{\partial \phi_2(y)}
   \nonumber\\
&&
G_{12}^0(x,y)\,(\stackrel{\leftarrow}{\square}_y+m^2)\,\phi_1(y)=-i\delta(x-y)\,\phi_1(y)\; .
\end{eqnarray}
By subtracting these equations and integrating over $y\in\Omega$, we
obtain
\begin{eqnarray}
\phi_1(x)=
  i\int_\Omega \rmd^4y\;G_{12}^0(x,y)\,\frac{\partial{\bs L}_{\rm int}(\phi_1,\phi_2)}{\partial \phi_2(y)}
  -i\int_\Omega \rmd^4y\;G_{12}^0(x,y)\,\stackrel{\leftrightarrow}{\square}_y\,
\phi_1(y)\; .
\end{eqnarray}
The second term of the right hand side is a total derivative
thanks to
\begin{equation}
A\,\stackrel{\leftrightarrow}{\square}\,B
=\partial_\mu\,\Big[A\,\stackrel{\leftrightarrow}{\partial}{}^\mu\,B\Big]\; .
\end{equation}
Therefore, this term can be rewritten as a surface integral extended
to the boundary of the domain $\Omega$. With reasonable assumptions on
the spatial localization of the source $J(x)$ that drives the field,
we may disregard the contribution from the boundary at spatial
infinity. The remaining boundaries are at the initial time $t_i$ and
final time $t_f$,
\begin{equation}
\phi_1(x)=i\int_\Omega \rmd^4y\;G_{12}^0(x,y)\,\frac{\partial{\bs L}_{\rm int}(\phi_1,\phi_2)}{\partial \phi_2(y)}
-i
\int \rmd^3\y\;\Big[
G_{12}^0(x,y)\,\stackrel{\leftrightarrow}{\partial}_{y_0}\,
\phi_1(y)
\Big]^{t_f}\; .
\label{eq:ode1}
\end{equation}
Note that the boundary term vanishes at the initial time $t_i$,
because $G_{12}^0$ is the retarded propagator.  Likewise, we obtain
the following equation for $\phi_2$:
\begin{eqnarray}
\phi_2(x)&=&i\int_\Omega \rmd^4y\;\Big\{
            G_{21}^0(x,y)\,\frac{\partial{\bs L}_{\rm int}(\phi_1,\phi_2)}{\partial \phi_1(y)}
             +
G_{22}^0(x,y)\,\frac{\partial{\bs L}_{\rm int}(\phi_1,\phi_2)}{\partial \phi_2(y)}
\Big\}\nonumber\\
&&\quad-i
\int \rmd^3\y\;\Big[
G_{21}^0(x,y)\,\stackrel{\leftrightarrow}{\partial}_{y_0}\,
\phi_2(y)
+
G_{22}^0(x,y)\,\stackrel{\leftrightarrow}{\partial}_{y_0}\,
\phi_1(y)
\Big]_{t_i}^{t_f}\; .
\nonumber\\
&&
\label{eq:ode2}
\end{eqnarray}
The boundary conditions at $t_i$ and $t_f$ are obtained by
comparing eqs.~(\ref{eq:int}) and (\ref{eq:ode1}-\ref{eq:ode2}).
At the final time $t_f$, the boundary condition is
\begin{equation}
\phi_1(t_f,\x)=0\quad,\qquad\partial_0\phi_1(t_f,\x)=i\,z(\x)\,{\cal O}'(\phi_2(t_f,\x))\; .
    \label{eq:bc-final}
\end{equation}
At the initial time $t_i$, we must have
\begin{equation}
\int_{y^0=t_i} \rmd^3\y\;\Big[
G_{21}^0(x,y)\,\stackrel{\leftrightarrow}{\partial}_{y_0}\,
\phi_2(y)
+
G_{22}^0(x,y)\,\stackrel{\leftrightarrow}{\partial}_{y_0}\,
\phi_1(y)
\Big]=0\; .
\end{equation}
Some simple manipulations lead to the following equivalent form
\begin{eqnarray}
  &&
     \int_{y^0=t_i} \rmd^3\y\;
G_{-+}^0(x,y)\,\stackrel{\leftrightarrow}{\partial}_{y_0}\,
\left(\phi_2(y)+\tfrac{1}{2}\phi_1(y)\right)
\nonumber\\
=&&
\int_{y^0=t_i} \rmd^3\y\;
G_{+-}^0(x,y)\,\stackrel{\leftrightarrow}{\partial}_{y_0}\,
\left(\phi_2(y)-\tfrac{1}{2}\phi_1(y)\right)
    =0\; .
    \label{eq:bc-init}
\end{eqnarray}
From the explicit form of the propagators $G_{+-}^0$ and $G_{-+}^0$
(see eqs.~(\ref{eq:props-vac})), we see that, at the initial time, the
combination $\phi_2+\tfrac{1}{2}\phi_1$ has no positive frequency
components, and the combination $\phi_2-\tfrac{1}{2}\phi_1$ has no
negative frequency components. An equivalent way to state this
boundary condition is in terms of the Fourier coefficients of the
fields $\phi_{1,2}$. Let us decompose them at the time $t_i$ as
follows,
\begin{equation}
  \phi_{1,2}(t_i,\x)\equiv \int_\k\Big\{
  {\wt{\bs\phi}}_{1,2}^{(+)}(\k)\,e^{-ik\cdot x}
  +
  {\wt{\bs\phi}}_{1,2}^{(-)}(\k)\,e^{+ik\cdot x}
  \Big\}\; .
\end{equation}
In terms of the coefficients introduced in this decomposition, the
boundary conditions at the initial time read:
\begin{equation}
  {\wt{\bs\phi}}_{2}^{(+)}(\k)=-\frac{1}{2}\,{\wt{\bs\phi}}_{1}^{(+)}(\k)\;,\quad
  {\wt{\bs\phi}}_{2}^{(-)}(\k)=\frac{1}{2}\,{\wt{\bs\phi}}_{1}^{(-)}(\k)\; .
  \label{eq:bc-ti-ii}
\end{equation}
The equations of motion (\ref{eq:eoms}), accompanied by the boundary
conditions (\ref{eq:bc-final}) and (\ref{eq:bc-ti-ii}), are equivalent
to the formulation of \cite{Weinberg:2008mc} (this reference uses the
fields $\phi_{\pm}$ of the in-in formalism, instead of the fields
$\phi_{1,2}$ of the retarded-advanced representation that we are using
here).

\section{Expansion of the solution in powers of $z(\x)$}
\label{sec:expansion}
\subsection{Setup of the expansion}
In the tree level approximation, the fields $\phi_{1,2}$ obey
classical equations of motion and satisfy coupled boundary conditions
(both at the initial and final times), a problem which is usually
extremely hard to solve, even numerically.  Moreover, one should keep
in mind that the $z$-dependence of the solutions arises entirely from
the boundary condition at the final time.

A first approach, that we shall pursue in this section, it to expand
the solutions in powers of $z$, by writing them as follows:
\begin{eqnarray}
  \phi_1(x)
  &\equiv&
           \phi^{(0)}_1(x)
           +\int \rmd^3\x_1 \;z(\x_1)\;\phi^{(1)}_1(x;\x_1)
           \nonumber\\
  &&\qquad\quad
           +\frac{1}{2!}\int \rmd^3\x_1 \rmd^3\x_2
           \;z(\x_1)z(\x_2)\;\phi^{(2)}_1(x;\x_1,\x_2)
     +\cdots
     \nonumber\\
  \phi_2(x)
  &\equiv&
           \phi^{(0)}_2(x)
           +\int \rmd^3\x_1 \;z(\x_1)\;\phi^{(1)}_2(x;\x_1)
           \nonumber\\
  &&\qquad\quad
           +\frac{1}{2!}\int \rmd^3\x_1 \rmd^3\x_2
           \;z(\x_1)z(\x_2)\;\phi^{(2)}_2(x;\x_1,\x_2)
     +\cdots
     \label{eq:expansion-z}
\end{eqnarray}
By construction, the coefficients of this expansion are symmetric
functions of the $\x_i$'s, and they are nothing but the functional
derivatives of $\phi_{1,2}(x)$ with respect to $z$,
\begin{equation}
  \phi_{1,2}^{(n)}(x;\x_1\cdots \x_n)=
  \left.\frac{\delta^n \phi_{1,2}(x)}{\delta z(\x_1)\cdots \delta z(\x_n)}\right|_{z=0}\; .
\end{equation}
The $n$-point correlation function is obtained by starting from the
first derivative of $\ln{\cal F}$, equal to ${\cal O}(\phi_2)$, and by
differentiating it $n-1$ times with respect to $z$. Using {\sl Fa\`a
  di Bruno}'s formula, this leads to
\begin{equation}
  \big<{\cal O}(x_1){\cal O}(x_2)\cdots{\cal O}(x_n)\big>_{\rm c}
  =\!\!\!\!
  \sum_{\pi\in\Pi(\{2\cdots n\})}
  \!\!\!\!
  {\cal O}^{(|\pi|)}(\phi_2^{(0)}(x_1))
  \prod_{\sigma\in \pi}
  \phi_2^{(|\sigma|)}(x_1;\{\x_i\}_{i\in\sigma})\; ,
  \label{eq:fdB}
\end{equation}
where the notations used in this equation are the following:
\begin{itemize}
\item $\Pi(\{2\cdots n\})$ is the set of the partitions of
  $\{2\cdots n\}$,
\item $\pi$ denotes one of these partitions, and $|\pi|$ is its
  cardinal, i.e. the number of blocks into which $\{2\cdots n\}$ is
  partitioned,
\item $\sigma$ denotes a block in the partition $\pi$, and $|\sigma|$
  the number of elements in this block.
\end{itemize}
From eq.~(\ref{eq:fdB}), it is clear that all the correlation
functions can be obtained from the coefficients in the expansion of
$\phi_2$ in powers of $z$.

The coefficients $\phi_{1,2}^{(n)}$ can be determined iteratively as
follows. By inserting eqs.~(\ref{eq:expansion-z}) in the equations of
motion (\ref{eq:eoms}) (or eqs.~(\ref{eq:eoms-phi4}) in the specific
case of a $\phi^4$ interaction), and by using the fact that the
functions $1$, $z(\x_1)$, $z(\x_1)z(\x_2),\cdots$ are linearly
independent, we obtain a system of coupled equations for the
coefficients. In a similar fashion, we obtain boundary conditions at
the initial and final time for the coefficients. These equations form
a ``triangular'' hierarchy, starting with equations that involve only
$\phi_{1,2}^{(0)}$, and where the equations for $\phi_{1,2}^{(n)}$
only contain the previously obtained $\phi_{1,2}^{(n')}$ with
$n'<n$. In the rest of this section, we solve the first two orders of
this hierarchy, in order to obtain the $1$-point and $2$-point
correlation functions.

\subsection{Order zero}
The order $0$ in $z$ is obtained by setting $z\equiv 0$ in the
equations of motion (\ref{eq:eoms}) and in the boundary conditions
(\ref{eq:bc-final}-\ref{eq:bc-init}). This leads to considerable
simplifications. Firstly, the boundary condition at $t_f$ for the
coefficient $\phi_1^{(0)}$ reads
\begin{equation}
  \phi_1^{(0)}(t_f,\x)=0\quad,\qquad \partial_0\phi_1^{(0)}(t_f,\x)=0\; .
\end{equation}
From the equation of motion for $\phi_1^{(0)}$, we conclude that
$\phi_1^{(0)}(x)$ is identically zero in the entire domain $\Omega$,
\begin{equation}
\forall x\in\Omega\;,\quad\phi_1^{(0)}(x)=0\; .
\end{equation}
Then, the boundary conditions at $t_i$ tell us that
\begin{equation}
\phi_2^{(0)}(t_i,\x)=0\quad,\qquad \partial_0 \phi_2^{(0)}(t_i,\x)=0\; ,
\end{equation}
and since $\phi_1^{(0)}\equiv 0$ the equation of motion for $\phi_2^{(0)}$
simplifies into
\begin{equation}
  (\square_x+m^2)\,\phi_2^{(0)}(x)={\cal L}_{\rm int}'(\phi_2^{(0)}(x))\; ,
  \label{eq:class-eom}
\end{equation}
i.e. the usual Euler-Lagrange equation for the theory under
consideration.  Therefore, the coefficient $\phi_1^{(0)}$ is zero, and
$\phi_2^{(0)}$ is the solution of the classical equation of motion
that vanishes (as well as its first time derivative) at the initial
time.  The expectation value of the observable at tree level is simply
given by
\begin{equation}
{\cal C}_{\{1\}}={\cal O}(\phi_2^{(0)}(x_1))\; .
\end{equation}
Since this object will appear repeatedly in the following, let us
introduce a simple graphical representation:\setbox1\hbox to
0.4cm{\resizebox*{0.4cm}{!}{\includegraphics{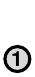}}} \setbox2\hbox to
6cm{\resizebox*{6cm}{!}{\includegraphics{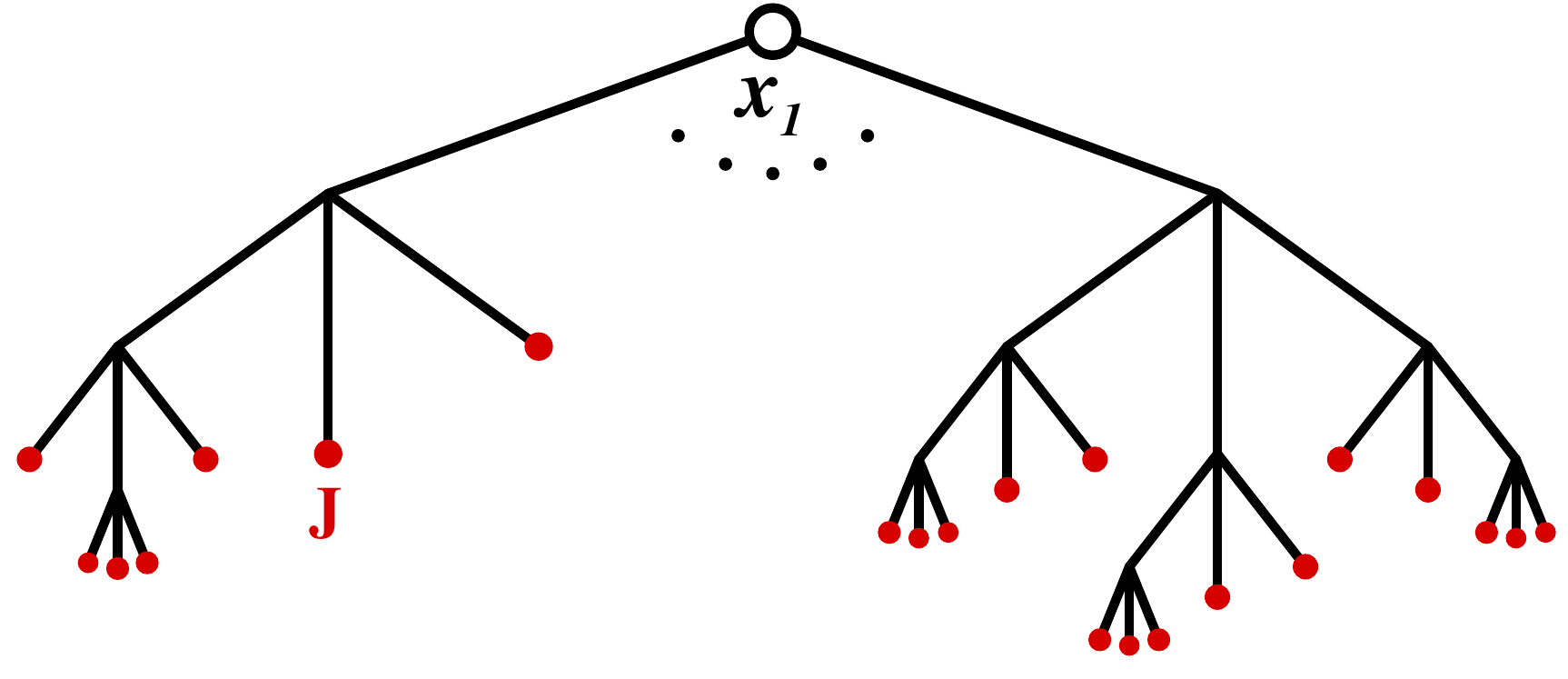}}}
\begin{equation}
{\cal C}_{\{1\}}\equiv
\raise -1.1mm\box1\;=\;\sum_{\rm trees}\raise -13mm\box2\; .
\label{eq:trees-1}
\end{equation}
The shaded blob in fact encapsulates an infinite series of tree
diagrams, a glimpse of which is given by the second equality.

\subsection{Order one}
\label{sec:2der}
\subsubsection{Equations of motion and boundary conditions}
Let us now consider the next order in $z$. Firstly, we obtain the
following two equations of motion,
\begin{eqnarray}
  &&\Big[\square+m^2-{\cal L}''_{\rm int}\big(\phi_2^{(0)}\big)\Big]\,\phi_1^{(1)}=0
     \nonumber\\
  &&\Big[\square+m^2-{\cal L}''_{\rm int}\big(\phi_2^{(0)}\big)\Big]\,\phi_2^{(1)}=0\; ,
     \label{eq:eoms-1}
\end{eqnarray}
where we have used the fact that $\phi_1^{(0)}=0$ to discard several
terms. Thus, both $\phi_1^{(1)}$ and $\phi_2^{(1)}$ obey the
linearized classical equation of motion about the $\phi_2^{(0)}$
background. The boundary condition at the final time reads
\begin{equation}
  \phi_1^{(1)}(t_f,\x;\x_1)=0\quad,\qquad
  \partial_0 \phi_1^{(1)}(t_f,\x;\x_1)=i\,\delta(\x-\x_1)\,{\cal O}'(\phi_2^{(0)}(t_f,\x_1))\; ,
  \label{eq:bc-final-1}
\end{equation}
while those at the initial time are
\begin{eqnarray}
  &&
     \int_{y^0=t_i} \rmd^3\y\;
G_{-+}^0(x,y)\,\stackrel{\leftrightarrow}{\partial}_{y_0}\,
\left(\phi_2^{(1)}(y;\x_1)+\tfrac{1}{2}\phi_1^{(1)}(y;\x_1)\right)
\nonumber\\
=&&
\int_{y^0=t_i} \rmd^3\y\;
G_{+-}^0(x,y)\,\stackrel{\leftrightarrow}{\partial}_{y_0}\,
\left(\phi_2^{(1)}(y;\x_1)-\tfrac{1}{2}\phi_1^{(1)}(y;\x_1)\right)
    =0\; ,
    \label{eq:bc-init-1}
\end{eqnarray}

\subsubsection{Solution in terms of mode functions}
\label{sec:modfunc}
The solution of this problem can be constructed as follows. Firstly,
let us introduce a complete basis of solutions of the linear equation
of motion that appears in eq.~(\ref{eq:eoms-1}). A convenient choice
will be to choose the functions of this basis such that they coincide
with plane waves at the initial time,
\begin{eqnarray}
&&
\Big[\square_x+m^2-{\cal L}''_{\rm int}(\phi_2^{(0)}(x))\Big]\,a_{\pm\k}(x)=0
\nonumber\\
&&\lim_{x^0\to t_i} a_{\pm\k}(x) = e^{\mp i k\cdot x}\; .
   \label{eq:modfunc}
\end{eqnarray}
These functions are sometimes called {\sl mode functions}.  The
boundary condition at $x^0\to t_i$ means that both the value of the
function and that of its first time derivative coincide with those of
the indicated plane wave. Thus, the functions $a_{+\k}(x)$ contain
only positive frequency modes at $t_i$, and the functions $a_{-\k}$
have only negative frequencies\footnote{Naturally, these statements
  are only true at $t_i$, since their subsequent propagation over a
  time dependent background will  mix positive and negative
  frequencies.}. The solution for $\phi_1^{(1)}$ can be expressed as a
linear combination of these mode functions as follows,
\begin{equation}
  \phi_1^{(1)}(x;\x_1)
  =\underbrace{\int_\k
  \Big\{a_{-\k}(x)a_{+\k}(t_f,\x_1)-a_{+\k}(x)a_{-\k}(t_f,\x_1)\Big\}}_{i\,G_{_A}(x,x_1)}\;{\cal O}'(\phi_2^{(0)}(t_f,\x_1))
  \; .
  \label{eq:phi11}
\end{equation}
That this function satisfies the required equation of motion is
obvious from the first of eqs.~(\ref{eq:modfunc}). The boundary
conditions at the final time follow from the properties
(\ref{eq:modes-ret}) of the mode functions that are derived in the
appendix \ref{app:modes} (up to a factor $i$, the underlined
combination of mode functions is the advanced propagator dressed by
the background field $\phi_2^{(0)}$). The boundary conditions at the
initial time tell us that the positive frequency content of
$\phi_2^{(1)}$ at $t_i$ is minus one-half that of $\phi_1^{(1)}$, and
its negative frequency content is plus one-half that of
$\phi_1^{(1)}$. Since the mode functions $a_{+\k}(x)$ and $a_{-\k}(x)$
have been defined in such a way that they contain only positive or
negative frequencies at $t_i$, respectively, it is immediate to write
the solution for $\phi_2^{(1)}$:
\begin{equation}
  \phi_2^{(1)}(x;\x_1)
  =\underbrace{\frac{1}{2}\int_\k
  \Big\{a_{-\k}(x)a_{+\k}(t_f,\x_1)+a_{+\k}(x)a_{-\k}(t_f,\x_1)\Big\}}_{G_{22}}\;{\cal O}'(\phi_2^{(0)}(t_f,\x_1))
  \; .
  \label{eq:phi21}
\end{equation}
Note that the underlined combination, that we denote $G_{22}$, is
nothing but the symmetric $22$ propagator dressed by the background
field $\phi_2^{(0)}$. The $2$-point correlation function is then given
by
\begin{eqnarray}
  {\cal C}_{\{12\}}=\big<{\cal O}(x_1){\cal O}(x_2)\big>
  &=&{\cal O}'(\phi_2^{(0)}(t_f,\x_2))\;\phi_2^{(1)}(t_f,\x_2)\nonumber\\
  &=&{\cal O}'(\phi_2^{(0)}(t_f,\x_2))\;G_{22}(x_1,x_2)\;{\cal O}'(\phi_2^{(0)}(t_f,\x_1))\; .
      \nonumber\\
  &&
     \label{eq:corr2}
\end{eqnarray}
Note that, although the final expression for the correlation function
is symmetric (as expected since the two observables with a space-like
separation commute), the intermediate calculations break this manifest
symmetry. Eq.~(\ref{eq:corr2}) generalizes to the non-linear strong
field regime the {\sl ``equation of motion method''} (see
\cite{Chen:2015dga} for instance) that has been devised for the
calculation of 2-point correlations of cosmological density
fluctuations.

\subsubsection{Expression in terms of derivatives w.r.t. the initial field}
Eq.~(\ref{eq:corr2}) can also be rewritten in a different way, that we
will generalize to the case of the $n$-point correlation in the strong
field approximation. Firstly, let us generalize the classical field
$\phi_2^{(0)}(x)$ (that has null initial conditions) into a classical
field $\Phi$ obeying the same equation of motion but with generic
initial condition $\Phi_{\rm ini}$ at $t_i$. Such a field obeys the
following Green's formula,
\begin{eqnarray}
\Phi(x)&=&
i\int_\Omega \rmd^4 y\; G_{_R}^0(x,y)\;{\cal L}'_{\rm int}(\Phi(y))
\nonumber\\
&&\qquad +i\int_{t_i} \rmd^3\y\; G_{_R}^0(x,y)\,
\stackrel{\leftrightarrow}{\partial}_{y^0}\,\Phi_{\rm ini}(y)\; ,
\end{eqnarray}
where $G_{_R}^0\equiv G_{++}^0-G_{+-}^0$ is the bare retarded
propagator. Note that, by definition
\begin{equation}
\phi_2^{(0)}=\left.\Phi\right|_{\Phi_{\rm ini}\equiv 0}\; .
\end{equation}
Then, let us write a similar integral representation for
the mode functions $a_{\pm\k}(x)$ introduced above,
\begin{eqnarray}
a_{\pm\k}(x)&=&
i\int_\Omega \rmd^4 y\; G_{_R}^0(x,y)\;{\cal L}''_{\rm int}(\phi_2^{(0)}(y))\;
a_{\pm\k}(y)
\nonumber\\
&&\qquad +i\int_{t_i} \rmd^3\y\; G_{_R}^0(x,y)\,
\stackrel{\leftrightarrow}{\partial}_{y^0}\,e^{\mp i k\cdot y}\; ,
\end{eqnarray}
By comparing these two integral representations, we see that we can
write the following formal relationship between $\Phi$ and
$a_{\pm \k}$:
\begin{equation}
a_{\pm\k}(x)=\left.\Bigg[\int_{t_i} \rmd^3\y\;
e^{\mp i k\cdot y}\frac{\delta}{\delta\Phi_{\rm ini}(y)}\Bigg]\,\Phi(x)\right|_{\Phi_{\rm ini}\equiv 0}\; .
\end{equation}
In other words, the mode function $a_{\pm\k}(x)$ can be viewed as the
functional derivative of the retarded classical field $\Phi$ with
respect to its initial condition, reweighted by the plane wave
$\exp(\mp i k\cdot x)$. This relationship implies that
\begin{equation}
  a_{\pm\k}(x)\,{\cal O}'(\phi_2^{(0)}(x))=
    \Bigg[\underbrace{\int_{t_i} \rmd^3\y\;
e^{\mp i k\cdot y}\frac{\delta}{\delta\Phi_{\rm ini}(y)}}_{{\mathbbm T}_{\pm\k}}\Bigg]\;{\cal O}(\Phi(x))\Bigg|_{\Phi_{\rm ini}\equiv 0}\; .
\end{equation}
Thus, the tree level $2$-point correlation function can be written as
\begin{eqnarray}
{\cal C}_{\{12\}}
&=&
\frac{1}{2}\int_\k
\Bigg\{
\Big[{\mathbbm T}_{+\k}\;{\cal O}(\Phi(x_1))\Big]
\Big[{\mathbbm T}_{-\k}\;{\cal O}(\Phi(x_2))\Big]
\nonumber\\
&&\underbrace{\qquad+
\Big[{\mathbbm T}_{-\k}\;{\cal O}(\Phi(x_1))\Big]
\Big[{\mathbbm T}_{+\k}\;{\cal O}(\Phi(x_2))\Big]
\Bigg\}\Bigg|_{\Phi_{\rm ini}\equiv 0}}_{{\cal O}(\Phi(x_1))\;\otimes\; {\cal O}(\Phi(x_2))}\; .
\end{eqnarray}
The $\otimes$ product\footnote{From its definition,
\begin{equation*}
  A\otimes B \equiv \frac{1}{2}\int_\k A\;\Big[
  \stackrel{\leftarrow}{\mathbbm T}_{+\k}
  \stackrel{\rightarrow}{\mathbbm T}_{-\k}
  +
  \stackrel{\leftarrow}{\mathbbm T}_{-\k}
  \stackrel{\rightarrow}{\mathbbm T}_{+\k}\Big]\;B\; ,
\end{equation*}
the $\otimes$ product is symmetric, $A\otimes B=B\otimes A$.}
introduced here is just a compact notation for the symmetrized
integration over the momentum $\k$. In this formulation, the $2$-point
correlation function is obtained by linking two copies of the
observable (at the points $x_1$ and $x_2$), with a ``link operator''
made of two derivatives with respect to the initial condition of the
classical field $\Phi$, one acting on each end of the link. We will
represent diagrammatically this link operation as follows:
\setbox1\hbox to 15mm{\resizebox*{15mm}{!}{\includegraphics{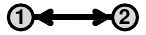}}}
\begin{eqnarray}
{\cal C}_{\{12\}}
\equiv\raise -1.2mm\box1\; .
\label{eq:C12}
\end{eqnarray}
It is important to keep in mind that this compact representation is
not a Feynman diagram, but rather a shorthand for an infinite series
of tree Feynman diagrams, made of two copies of the graphs that appear
in eq.~(\ref{eq:trees-1}) connected by a bare $G_{22}^0$ propagator in
all the possible ways: \setbox1\hbox to
80mm{\resizebox*{80mm}{!}{\includegraphics{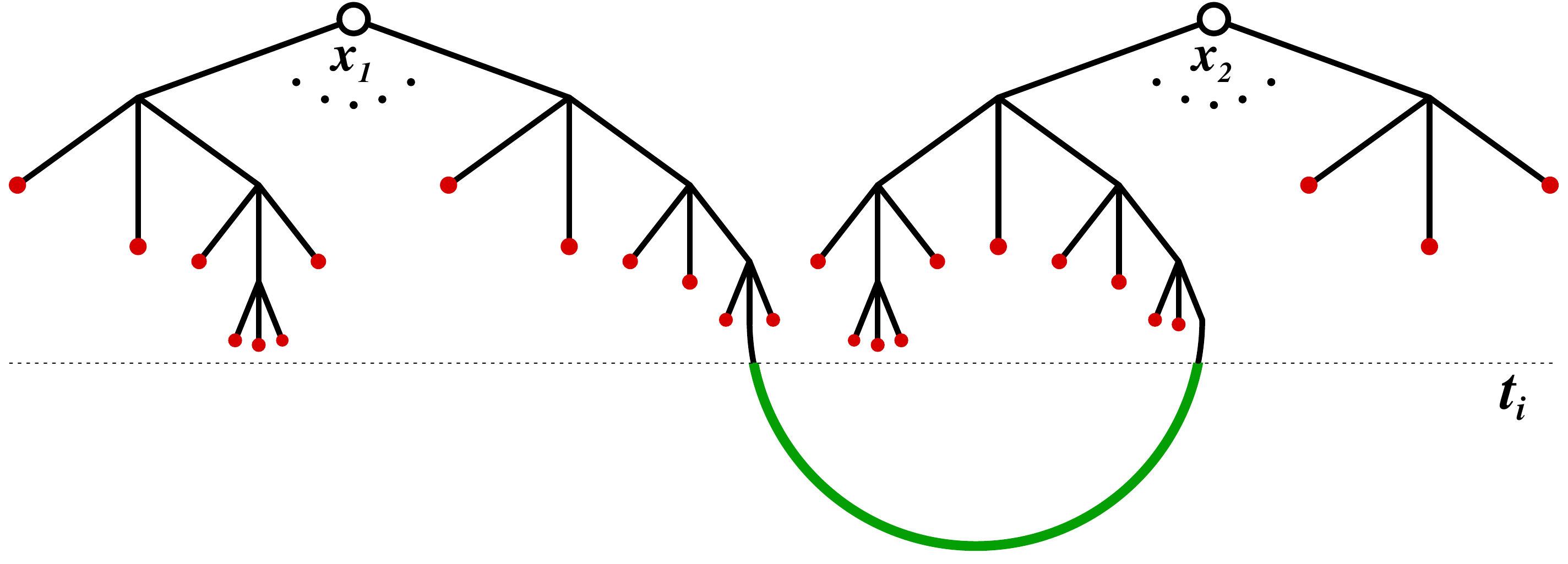}}}
\begin{equation}
  {\cal C}_{\{12\}}=\sum_{\rm trees}\;\;\raise -15mm\box1
  \label{eq:trees-2}
\end{equation}
This representation of the tree-level $2$-point correlation function
was already contained in the NLO result obtained in
\cite{Gelis:2008ad}, but the retarded-advanced basis used in the
present paper simplifies considerably its derivation.

The main result of this section is that all these graphs can be
obtained from the $1$-point function evaluated for a classical field
$\Phi$ with generic initial condition $\Phi_{\rm ini}$, by taking
functional derivatives with respect to this initial condition.  The
``link'' that connects the two handles created by these derivatives
encodes the zero-point fluctuations in the initial vacuum state. For
higher-point correlation functions at tree level, this result is not
true in general. Starting with the $3$-point correlation function (see
the section \ref{sec:beyond}), there are additional contributions that
cannot be expressed in terms of functional derivatives with respect to
the initial classical field $\Phi_{\rm ini}$. However, as we shall see
in the next section, this remains valid in the {strong field
  approximation}.

\section{Correlations in the strong field regime}
\label{sec:strong}
\subsection{Strong field approximation}
Until now, our counting was based on the fact that a large external
source $J$ leads to large fields $\phi_\pm$, but no approximation was
made in the calculation of the $1$-point and $2$-point correlation
functions at tree level in the previous section. Instead of pursuing
the very cumbersome expansion in powers of $z$ that we have used so
far, we consider in this section an approximation that allows a
formal solution to all orders in $z$. Here, we give only a very
sketchy motivation for this approximation, and a lengthier discussion
of its validity will be provided later in this section (after we have
derived expressions for the fields $\phi_1$ and $\phi_2$).

\begin{figure}[htbp]
  \begin{center}
    \resizebox*{11cm}{!}{\includegraphics{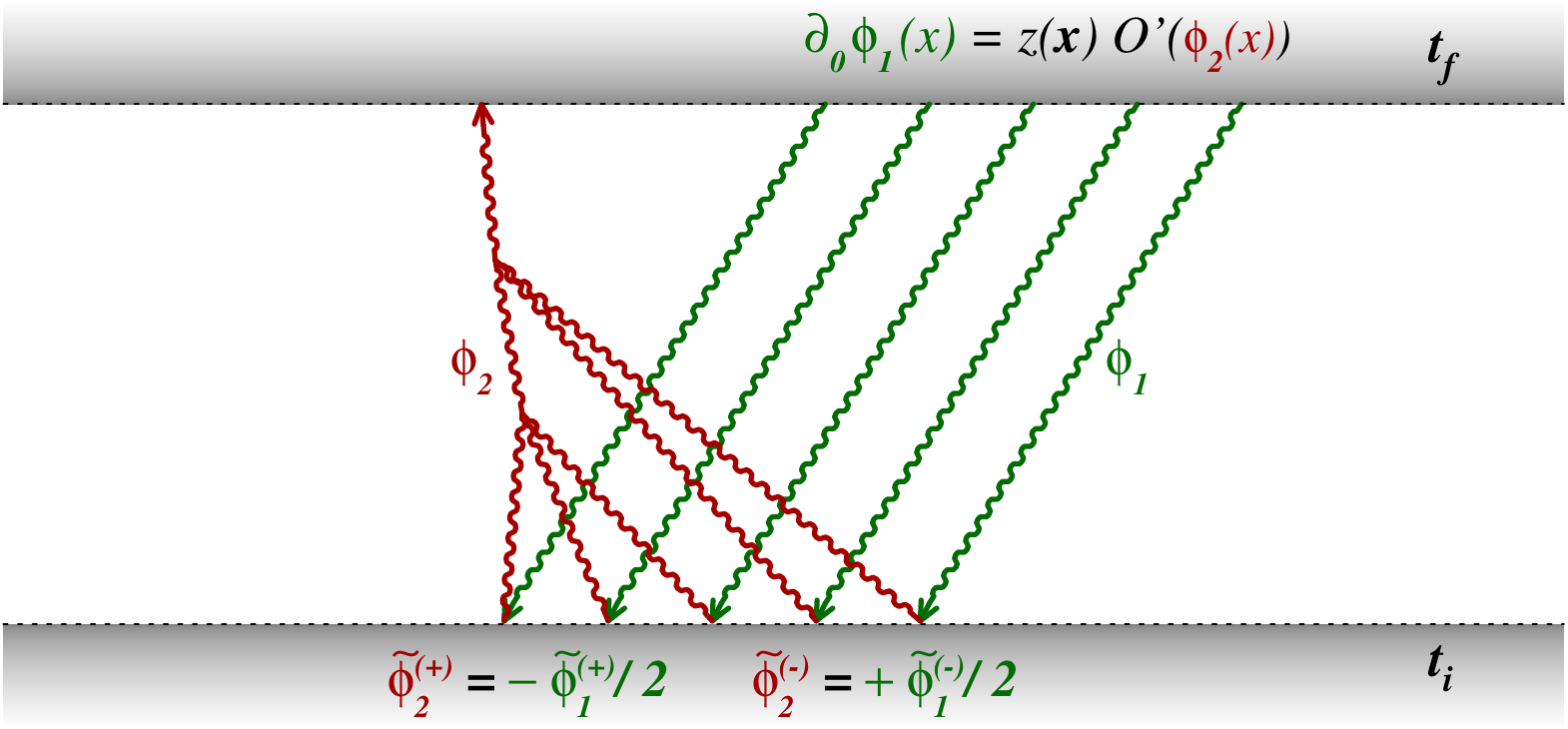}}
  \end{center}
  \caption{\label{fig:phi12} Relationship between the fields $\phi_1$
    and $\phi_2$ in the strong field approximation. }
\end{figure}

Let us first recall that the fields $\phi_+$ and $\phi_-$ represent,
respectively, the space-time evolution of the field in amplitudes and
in conjugate amplitudes. The fact that they are distinct leads to
interferences when squaring amplitudes, a quantum effect
controlled by $\hbar$. Consequently, we may expect the difference
$\phi_1\equiv \phi_+-\phi_-$ to be small compared to $\phi_\pm$
themselves, i.e.
\begin{equation}
  \phi_1\ll \phi_2\; .
  \label{eq:SFA}
\end{equation}
In this regime, that we will call the {\sl strong field approximation}
(SFA), we can approximate the equations of motion\footnote{The strong
  field approximation (\ref{eq:SFA}) is known to be non-renormalizable
  \cite{Epelbaum:2014yja}. However, this is not an issue in the
  present paper, since we are considering only tree-level
  contributions.} (\ref{eq:eoms}) by keeping only the lowest order in
$\phi_1$. This amounts to keeping only the terms linear in $\phi_1$ in
eq.~(\ref{eq:LL}) (in the case of a $\phi^4$ theory, it means dropping
the $\phi_1^3\phi_2$ term in eq.~(\ref{eq:LL-phi4})). In the
approximation, they read
\begin{eqnarray}
  &&\Big[\square+m^2-{\cal L}''_{\rm int}(\phi_2)\Big]\,\phi_1=0\; ,\nonumber\\
  &&(\square+m^2)\,\phi_2-{\cal L}'_{\rm int}(\phi_2)=0\; ,
     \label{eq:eoms-sfr}
\end{eqnarray}
while the boundary conditions are still given by (\ref{eq:bc-final})
and (\ref{eq:bc-init}). The problem one must now solve is illustrated in
the figure \ref{fig:phi12}.
The field $\phi_1$ obeys a linear equation of motion (dressed by the
field $\phi_2$, although this aspect is not visible in the figure),
with an advanced boundary condition that depends on $\phi_2$. In
parallel, the field $\phi_2$ obeys a non-linear equation of motion,
with a retarded boundary condition that depends on $\phi_1$. As we
shall show in the next subsection, this tightly constrained problem
admits a formal solution, valid to all orders in the function $z$, in
the form of an implicit functional equation for the first derivative of $\ln{\cal F}[z]$.

\subsection{Formal solution}
The equation of motion for $\phi_1$ (first of
eqs.~(\ref{eq:eoms-sfr})) is formally identical to the first of
eqs.~(\ref{eq:eoms-1}). We can therefore mimic eq.~(\ref{eq:phi11})
and write directly $\phi_1$  as follows:
\begin{equation}
  \phi_1(x)
  =\int_\k\int \rmd^3\u\;
  \Big\{a_{-\k}(x)a_{+\k}(t_f,\u)-a_{+\k}(x)a_{-\k}(t_f,\u)\Big\}\;z(\u)\;{\cal O}'(\phi_2(t_f,\u))
  \; ,
  \label{eq:phi1}
\end{equation}
where the mode functions $a_{\pm\k}$ should now be defined with
$\phi_2$ as the background, rather than $\phi_2^{(0)}$. To obtain
eq.~(\ref{eq:phi1}), it was crucial to have a linear equation of
motion for $\phi_1$, a consequence of the strong field
approximation. The above equation formally defines $\phi_1(x)$ in the
bulk, $x\in\Omega$, in terms of the field $\phi_2$ at the final
time. It is important to note that this equation is valid to all
orders in $z$, contrary to the equations encountered in the expansion
in powers of $z$ that we have used in the previous section. Beside the
explicit factor $z(\u)$, the right hand side contains also an implicit
$z$ dependence (to all orders in $z$) in the field $\phi_2(t_f,\u)$
and in the mode functions $a_{\pm\k}$ (since they evolve on top of the
background $\phi_2$).

Then, using the boundary condition at the initial time, we obtain the
following expression for the field $\phi_2$ at $t_i$,
\begin{eqnarray}
  \phi_2(t_i,\y)
  &=&\frac{1}{2}\int_\k\int \rmd^3\u\;
     \Big\{e^{+ik\cdot y}\,a_{+\k}(t_f,\u)\nonumber\\
  &&\quad\qquad\qquad
     +e^{-ik\cdot y}\,a_{-\k}(t_f,\u)\Big\}\;z(\u)\;{\cal O}'(\phi_2(t_f,\u))
     \nonumber\\
  &=&\frac{1}{2}\int_\k\int \rmd^3\u\;z(\u)\;{\cal O}(\phi_2(t_f,\u))\;
      \Big\{
      \stackrel{\leftarrow}{\mathbbm T}_{+\k}\,e^{+ik\cdot y}
      +
      \stackrel{\leftarrow}{\mathbbm T}_{-\k}\,e^{-ik\cdot y}\Big\}\;
      \; ,\nonumber\\
  &&
  \label{eq:phi2-ti}
\end{eqnarray}
where the arrows indicate on which side the ${\mathbbm T}_{\pm\k}$
operators act.  This expression for the field $\phi_2$ at the time
$t_i$ can be used as initial condition for the first of
eqs.~(\ref{eq:eoms-sfr}).  The next step is to note that the field
$\phi_2(x)$ that satisfies this equation of motion, and has the
initial condition $\phi_2(t_i,\y)$ is formally given by
\begin{equation}
  \phi_2(x)=\exp\Bigg\{\int \rmd^3\y\;\phi_2(t_i,\y)\frac{\delta}{\delta\Phi_{\rm ini}(t_i,\y)}\Bigg\}\;\Phi(x)\Bigg|_{\Phi_{\rm ini}\equiv 0}\; .
  \label{eq:translation}
\end{equation}
This formula follows from the fact that the derivative
$\delta/\delta\Phi_{\rm ini}$ is the generator for shifts of the
initial condition of $\Phi$; its exponential is therefore the
corresponding translation operator.  The same formula applies also to
any function of the field, since the exponential operator shifts
$\Phi_{\rm ini}$ in any instance of $\Phi$ on its right. In
particular, we have
\begin{equation}
  {\cal O}\big(\phi_2(x)\big)
  =\exp\Bigg\{\int \rmd^3\y\;\phi_2(t_i,\y)\frac{\delta}{\delta\Phi_{\rm ini}(t_i,\y)}\Bigg\}\;{\cal O}\big(\Phi(x)\big)\Bigg|_{\Phi_{\rm ini}\equiv 0}\; .
  \label{eq:translation-1}
\end{equation}
Substituting $\phi_2(t_i,\y)$ by eq.~(\ref{eq:phi2-ti}) inside the
exponential, this leads to
\begin{eqnarray}
  {\cal O}\big(\phi_2(x)\big)
  &=&
      \exp\smash{\Bigg\{}
     \frac{1}{2}\int_\k\int \rmd^3\u\;z(\u)\,{\cal O}(\phi_2(t_f,\u))\nonumber\\
  &&\qquad\qquad
      \times\Big[
      \stackrel{\leftarrow}{\mathbbm T}_{+\k}
      \stackrel{\rightarrow}{\mathbbm T}_{-\k}
      +
      \stackrel{\leftarrow}{\mathbbm T}_{-\k}
      \stackrel{\rightarrow}{\mathbbm T}_{+\k}\Big]
      \smash{\Bigg\}}\;{\cal O}\big(\Phi(x)\big)\smash{\Bigg|}_{\Phi_{\rm ini}\equiv 0}
     \nonumber\\
  &&\nonumber\\
  &=&\exp\Bigg\{
      \int \rmd^3\u\;z(\u)\,{\cal O}(\phi_2(t_f,\u))\;\otimes
      \Bigg\}\;{\cal O}\big(\Phi(x)\big)\Bigg|_{\Phi_{\rm ini}\equiv 0}\; .
\end{eqnarray}
Setting $x^0=t_f$ and denoting 
\begin{equation}
{\cal D}[\x_1;z]\equiv {\cal O}(\phi_2(t_f,\x_1))
\end{equation}
the first derivative of $\ln{\cal F}$, we see that it obeys the
following recursive formula
\begin{equation}
  {\cal D}[\x_1;z]
  =\exp\Bigg\{
  \int \rmd^3\u\;z(\u)\,{\cal D}[\u;z]\;\otimes
  \Bigg\}\;{\cal O}\big(\Phi(t_f,\x_1)\big)\Bigg|_{\Phi_{\rm ini}\equiv 0}\; .
  \label{eq:master-D}
\end{equation}

Let us make a technical remark about the scope of the
$\stackrel{\leftarrow}{\mathbbm T}_{\pm\k}$ derivative (contained in
the $\otimes$ product) that acts on the factor ${\cal D}[\u;z]$ inside
the exponential.
\begin{figure}[htbp]
  \begin{center}
    \resizebox*{11cm}{!}{\includegraphics{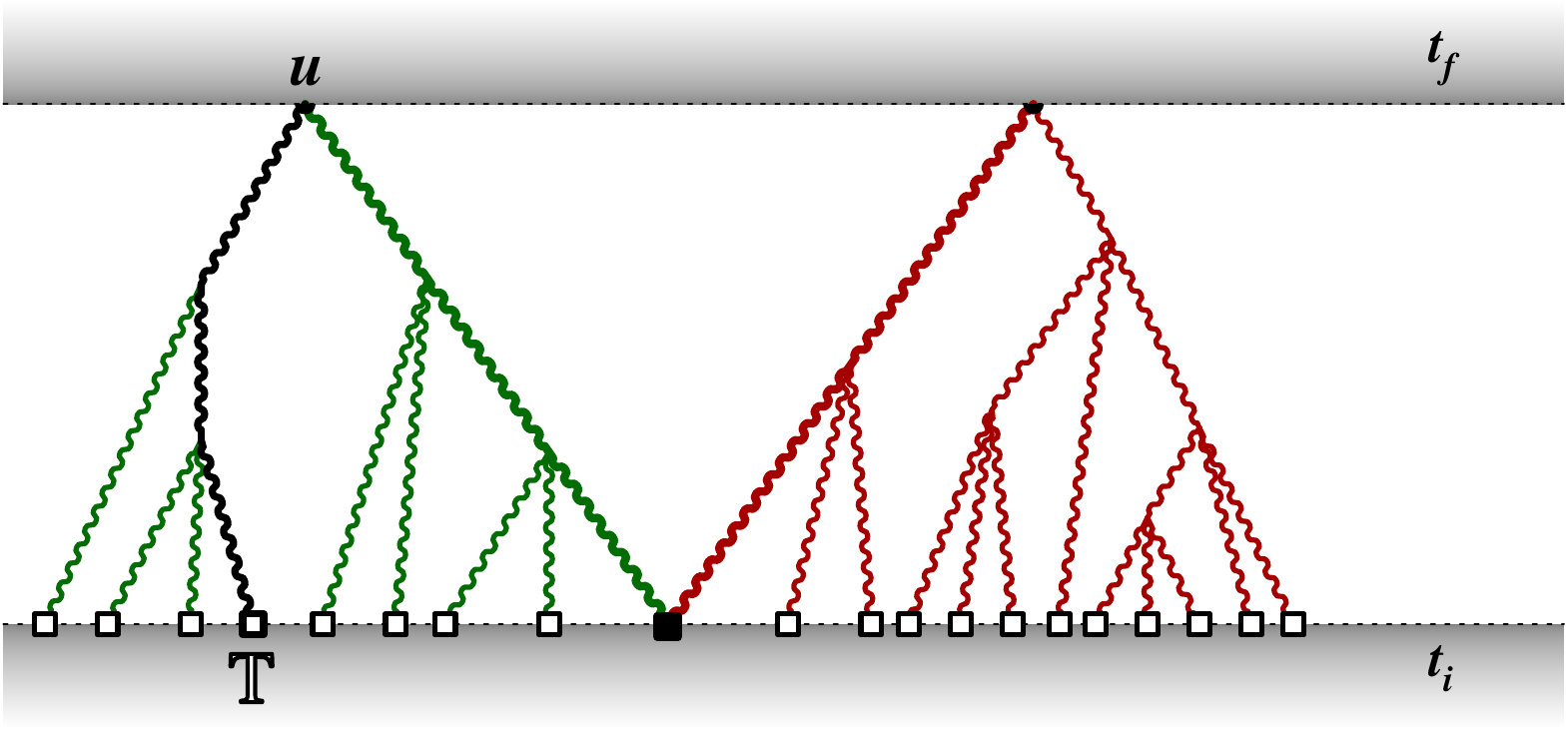}}
  \end{center}
  \caption{\label{fig:deri} Typical term in $D[\u;z]$ (here we have
    shown a term with two nodes), showing the tree dependence on the
    initial field $\Phi_{\rm ini}$ (represented by open squares on the
    initial time surface). The filled square represents the link
    between the two nodes, that involves the function of
    eq.~(\ref{eq:bareG22}).  The $\otimes$ product acting on $D[\u;z]$  in
    eq.~(\ref{eq:master-D}) acts only on the initial fields that
    connect to the point $\u$.}
\end{figure}
This derivative originates from the terms in
$a_{\pm\k}(t_f,\u)\,{\cal O}'(\phi_2(t_f,\u))$ in
eq.~(\ref{eq:phi2-ti}). From this origin, it is clear that 
$\stackrel{\leftarrow}{\mathbbm T}_{\pm\k}$ must establish a
link between a point on the initial surface and the point $\u$ on the
final surface, as illustrated in the figure \ref{fig:deri}, where the
black line is the function $a_{\pm \k}(t_f,\u)$). As we shall see in
the next section, the recursive expansion of eq.~(\ref{eq:master-D})
leads to a tree structure where the nodes are observables
${\cal O}(\Phi(t_f,\x))$. The above discussion tells us that when
performing this expansion, the $\otimes$ product should not be
``distributed'' to all the nodes inside $D[\u;z]$, but applied only to the
node of coordinate $\u$.

\subsection{Realization of the strong field approximation}
\label{sec:insta}
Let us now return on the condition $ \phi_1 \ll \phi_2 $ , that was
used in the derivation of eq.~(\ref{eq:master-D}), in order to see a
posteriori when it is satisfied. To that effect, we can use
eq.~(\ref{eq:phi1}) for $\phi_1$. For $\phi_2$, the initial condition
at $t_i$ is given by eq.~(\ref{eq:phi2-ti}). For the sake of this
discussion, it is sufficient to use a linearized solution for $\phi_2$
in the bulk, that reads
\begin{eqnarray}
  \phi_2(x)\Big|_{\rm lin}
  &=&
      \frac{1}{2}\int_\k\int \rmd^3\u\;
      \Big\{a_{-\k}(x)a_{+\k}(t_f,\u)+a_{+\k}(x)a_{-\k}(t_f,\u)\Big\}\nonumber\\
  &&\qquad\qquad\qquad\qquad\qquad\qquad
     \times \;z(\u)\;{\cal O}'(\phi_2(t_f,\u))
  \; ,
  \label{eq:phi2-lin}
\end{eqnarray}
First of all, a comparison between eqs.~(\ref{eq:phi1}) and
(\ref{eq:phi2-lin}) indicates that $\phi_1$ and $\phi_2$ have the same
order in the coupling constant $g$, since they are made of the same
building blocks (the only difference is the sign between the two terms
of the integrand, and an irrelevant overall factor
$\tfrac{1}{2}$).

However, a hierarchy between $\phi_1$ and $\phi_2$ arises dynamically
when the classical solutions of the field equation of motion
(\ref{eq:class-eom}) are unstable. Such instabilities are fairly
generic in several quantum field theories; in particular the scalar
field theory with a $\phi^4$ coupling that we are using as example
throughout this paper is known to have a parametric resonance
\cite{Greene:1997fu,Dusling:2010rm}.  Since the mode functions
$a_{\pm\k}$ are linearized perturbations on top of the classical field
$\phi_2$, an instability of the classical solution $\phi_2$ is
equivalent to the fact that some of the mode functions grow
exponentially with time, as $\exp(\mu(x^0-t_i))$ (where $\mu$ is the
Lyapunov exponent).  Thus, since eq.~(\ref{eq:phi2-lin}) is bilinear
in the mode functions, we expect that
\begin{equation}
\phi_2(x)\Big|_{\rm lin} \sim e^{\mu(x^0+t_f-2t_i)} \; .
\end{equation}
Estimating the magnitude of $\phi_1$ requires more care. Indeed, from
eqs.~(\ref{eq:modes-ret}) in the appendix \ref{app:modes},
antisymmetric combinations of the mode functions at equal times
remain of order $1$ even if individual mode functions grow
exponentially with time. Thus, at the final time, we have
\begin{equation}
\phi_1(t_f,\x)\sim 1\quad\mbox{and}\quad \frac{\phi_2(t_f,\x)}{\phi_1(t_f,\x)}\sim e^{2\mu(t_f-t_i)}\gg 1\; ,
\end{equation}
for sufficiently large $t_f-t_i$.

In order to estimate the ratio $\phi_2/\phi_1$ at intermediate times,
one may use the following reasoning. The antisymmetric combination of
mode functions that enters in eq.~(\ref{eq:phi1}) is the advanced
propagator $G_{_A}$ in the background $\phi_2$. This retarded
propagator may also be expressed in terms of a different set of mode
functions $b_{\pm\k}$ defined to be plane waves at the final time
$t_f$,
\begin{eqnarray}
&&
\Big[\square_x+m^2-{\cal L}''_{\rm int}(\phi_2(x))\Big]\,b_{\pm\k}(x)=0
\nonumber\\
&&\lim_{x^0\to t_f} b_{\pm\k}(x) = e^{\mp i k\cdot x}\; .
   \label{eq:modfunc-b}
\end{eqnarray}
In terms of these alternate mode functions, we also have
\begin{equation}
  \phi_1(x)
  =\int_\k\int \rmd^3\u\;
  \Big\{b_{-\k}(x)b_{+\k}(t_f,\u)-b_{+\k}(x)b_{-\k}(t_f,\u)\Big\}\;z(\u)\;{\cal O}'(\phi_2(t_f,\u))
  \; ,
  \label{eq:phi1-alt}
\end{equation}
In the presence of instabilities, these backward evolving mode
functions grow when $x^0$ decreases away from $t_f$, as
$\exp(\mu(t_f-x^0))$ (in this sketchy argument, the Lyapunov exponent
$\mu$ is assumed here to be the same for the forward and backward mode
functions). This implies
\begin{equation}
\phi_1(x)\sim e^{\mu(t_f-x^0)}\; ,
\end{equation}
and the following magnitude for the ratio $\phi_2/\phi_1$ at
intermediate times
\begin{equation}
\frac{\phi_2(x)}{\phi_1(x)}\sim \frac{e^{\mu(x^0+t_f-2t_i)}}{e^{\mu(t_f-x^0)}}\sim e^{2\mu(x^0-t_i)}\; .
\end{equation}
Thus, with instabilities and non-zero Lyapunov exponents, the strong
field approximation is generically satisfied thanks to the exponential
growth of perturbations over the background. In the appendix
\ref{app:mixed}, we will discuss another situation where this
approximation is also satisfied, namely when the initial state is a mixed
state with large occupation number.

\subsection{Expansion of eq.~(\ref{eq:master-D}) in powers of $z$}
Although eq.~(\ref{eq:master-D}) cannot be solved explicitly, it is
fairly easy to obtain a diagrammatic representation of its
solution. For this, let us introduce the following graphical
notations: \setbox1\hbox to
5mm{\resizebox*{5mm}{!}{\includegraphics{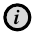}}} \setbox2\hbox to
5mm{\resizebox*{5mm}{!}{\includegraphics{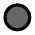}}} \setbox3\hbox to
21mm{\resizebox*{21mm}{!}{\includegraphics{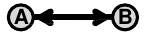}}}
\begin{equation}
\begin{aligned}
&&\raise -1.7mm\box1 \quad&\equiv&&\quad {\cal O}\big(\Phi(t_f,\x_i)\big)\; ,&\nonumber\\
&&\raise -1.7mm\box2 \quad&\equiv&&\quad \int \rmd^3\u\;z(\u)\;{\cal O}\big(\Phi(t_f,\u)\big)\; ,&\nonumber\\
&&\raise -1.7mm\box3 \quad&\equiv&&\quad A\otimes B\; .&
\end{aligned}
\end{equation}
\setbox1\hbox to 3.63mm{\resizebox*{3.63mm}{!}{\includegraphics{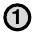}}} At the
order $0$ in $z$, we just need to set $z\equiv 0$ inside the
exponential, to obtain
\begin{equation}
{\cal D}^{(0)}[\x_1;z] = \raise -1mm\box1\; .
\end{equation}
Then, we proceed recursively. We insert the $0$-th order result in the
exponential, and expand to order $1$ in $z$, leading to the following result at order $1$:
\setbox2\hbox to 15mm{\resizebox*{15mm}{!}{\includegraphics{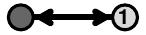}}}
\begin{equation}
{\cal D}^{(1)}[\x_1;z] = \raise -1mm\box2\; .
\end{equation}
\setbox1\hbox to 26mm{\resizebox*{26mm}{!}{\includegraphics{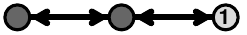}}}%
\setbox2\hbox to 26mm{\resizebox*{26mm}{!}{\includegraphics{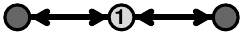}}}%
The next two iterations give:
\begin{equation}
{\cal D}^{(2)}[\x_1;z] = \raise -1mm\box1+\frac{1}{2!}\,\raise -1mm\box2\; ,
\end{equation}
and
\setbox1\hbox to 37mm{\resizebox*{37mm}{!}{\includegraphics{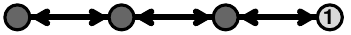}}}
\setbox2\hbox to 37mm{\resizebox*{37mm}{!}{\includegraphics{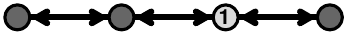}}}
\setbox3\hbox to 20mm{\resizebox*{20mm}{!}{\includegraphics{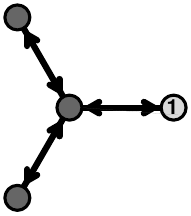}}}
\setbox4\hbox to 20mm{\resizebox*{20mm}{!}{\includegraphics{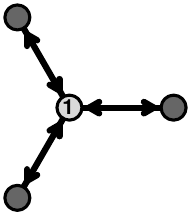}}}
\begin{eqnarray}
  {\cal D}^{(3)}[\x_1;z] &=& \raise -1mm\box1+\;\raise -1mm\box2\nonumber\\
                  &&\qquad\quad\vphantom{\Bigg[}+\frac{1}{2!}\raise -11mm\box3+\frac{1}{3!}\!\!\raise -11mm\box4\; .
\end{eqnarray}
These examples generalize to all orders in $z$: the functional
${\cal D}[\x_1;z]$ can be represented as the sum of all the rooted
trees (the root being the node carrying the fixed point $\x_1$) weighted by
the corresponding symmetry factor $1/{\cal S}(T)$: \setbox1\hbox to
37.5mm{\resizebox*{37.5mm}{!}{\includegraphics{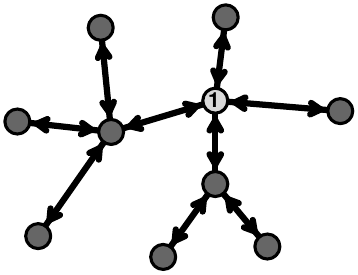}}}
\begin{equation}
\frac{\delta \ln{\cal F}[z]}{\delta z(\z_1)}={\cal D}[\x_1;z]=\sum_{\mbox{\scriptsize rooted}\atop\mbox{\scriptsize trees }T}\frac{1}{{\cal S}(T)}\;\;\raise -17mm\box1\; .
\end{equation}
Note that the sum of the weights for the trees with $n+1$ nodes (one
of them being the root node $\x_1$),
\begin{equation}
  w_n\equiv\sum_{\mbox{\scriptsize rooted}\atop\mbox{\scriptsize trees }T_{n+1}} \frac{1}{{\cal S}(T_{n+1})}\; ,
\end{equation}
is the $n$-th Taylor coefficient of the function $w(z)$,
\begin{equation}
w(z)\equiv \sum_{n=0}^{\infty} w_n\,z^n\; ,
\end{equation}
that satisfies the following identity
\begin{equation}
  w(z)=e^{z\,w(z)}\; .
  \label{eq:ww}
\end{equation}
This functional identity may be viewed as a structureless version of
eq.~(\ref{eq:master-D}), in which the function $z(\x)$ is replaced by
a scalar variable $z$. Eq.~(\ref{eq:ww}) leads to (see
\cite{flagolet}, pp. 127-128)
\begin{equation}
w_n=\frac{(n+1)^{n-1}}{n!}\; ,
\end{equation}
which is the number of trees with $n+1$ labeled nodes ({\sl Cayley's
  formula}) divided by the number of ways of reshuffling the $n$ nodes
that are integrated out.

\subsection{Correlation functions}
The $n$-point correlation function is obtained by differentiating this
expression $n-1$ times, with respect to $z(\x_2),\cdots,z(\x_n)$, and
by setting $z\equiv 0 $ afterwards. This selects all the trees with
$n$ distinct labeled nodes\footnote{Thus, permuting nodes in general
  yields a different tree.} (including the node at $\x_1$). Moreover,
since derivatives commute, these successive differentiations eliminate
the symmetry factors, leading to \setbox1\hbox to
37.5mm{\resizebox*{37.5mm}{!}{\includegraphics{tree_example2}}}
\begin{equation}
  \frac{\delta \ln{\cal F}[z]}{\delta z(\z_1)\cdots\delta z(\x_n)}\Bigg|_{z\equiv 0}={\cal C}_{\{1\cdots n\}}=\sum_{{\mbox{\scriptsize trees with }n}\atop{\mbox{\scriptsize labeled  nodes}}}\raise -17mm\box1\; .
  \label{eq:Cn-trees}
\end{equation}
The number of trees contributing to this sum is equal to
$n^{n-2}$. Eq.~(\ref{eq:Cn-trees}) tells us that, at tree level in the
strong field regime, all the $n$-point correlation functions are
entirely determined by the functional dependence of the solution of
the classical field equation of motion with respect to its initial
condition. Moreover, this formula provides a way to construct
explicitly the correlation functions in terms of functional
derivatives with respect to the initial field.

In this tree representation, the number of links reaching a node is
the number derivatives with respect to $\Phi_{\rm ini}$ that act on
the corresponding ${\cal O}(\Phi)$. When the observable is a composite
operator in terms of the fields of the theory, this includes a large
number of contributions. For instance, when ${\cal O}(\Phi(x))$ is
cubic in the field $\Phi(x)$, a node with three links would contain
the following terms: \setbox1\hbox to
8mm{\resizebox*{8mm}{!}{\includegraphics{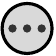}}} \setbox2\hbox to
14.5mm{\resizebox*{14.5mm}{!}{\includegraphics{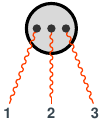}}} \setbox3\hbox
to 11.5mm{\resizebox*{11.5mm}{!}{\includegraphics{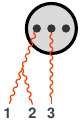}}}
\setbox4\hbox to
11.5mm{\resizebox*{11.5mm}{!}{\includegraphics{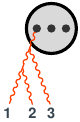}}} \setbox5\hbox
to 11.5mm{\resizebox*{11.5mm}{!}{\includegraphics{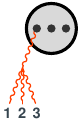}}}
\begin{equation}
  {\mathbbm T}_1{\mathbbm T}_2{\mathbbm T}_3\;\raise -3mm\box1
  \;\,
  =\raise -12.8mm\box2+\raise -12.5mm\box3\;\;+\raise -12.5mm\box4\;
  \;+\raise -12.5mm\box5\;\; ,
  \label{eq:tmp2}
\end{equation}
where the three dots inside the blob represent the three fields $\Phi$
it contains. Instead of the functional derivation of the
representation (\ref{eq:Cn-trees}) that we have performed in this
section, one could in principle have iterated on the $z$-expansion
introduced in the previous section (after simplifying the equations of
motion based on $\phi_1\ll \phi_2$). This approach produces a sum of
terms corresponding to the right hand side of eq.~(\ref{eq:tmp2})
that, after some hefty combinatorics, one may rewrite in the compact
form of the left hand side of eq.~(\ref{eq:tmp2}).

\begin{figure}[htbp]
  \begin{center}
    \resizebox*{11cm}{!}{\includegraphics{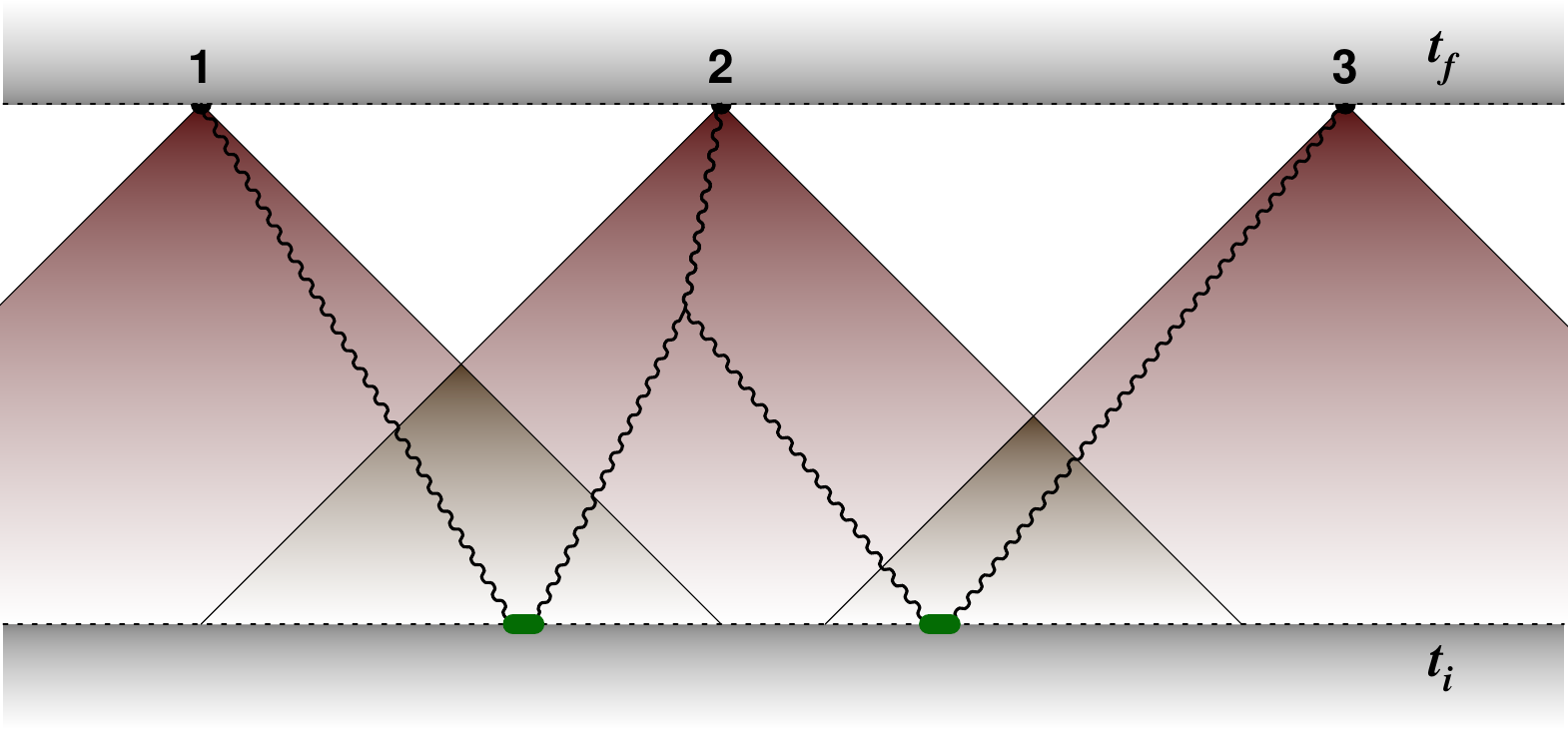}}
  \end{center}
  \caption{\label{fig:LC-sfr} Causal structure of the $3$-point
    correlation function in the strong field regime. }
\end{figure}

In the strong field approximation, the final state correlations are
entirely due to quantum fluctuations in the initial state, that are
encoded in the function $G_{22}^0(x,y)$. If the initial state is the
vacuum, as we have assumed in this paper, it reads
\begin{equation}
  G_{22}^0(x,y)=\int \frac{\rmd^3 \k}{(2\pi)^2 2E_\k}\;e^{i\k\cdot(\x-\y)}\; .
  \label{eq:bareG22}
\end{equation}
The support of this function is dominated by distances $|\x-\y|$
smaller than the Compton wavelength $m^{-1}$. Thus, in the tree
representation of eq.~(\ref{eq:Cn-trees}), a link between the points
$\x_i$ and $\x_j$ is nonzero provided that the past light-cones of
summits $x_i$ and $x_j$ overlap at the initial time (or at least
approach each other within distances $\lesssim m^{-1}$), as
illustrated in the figure \ref{fig:LC-sfr}.

\subsection{Generalizations}
Let us list here several extensions, for which the result of
eq.~(\ref{eq:Cn-trees}) would remain valid modulo some minor changes:
\begin{itemize}
\item Although we have assumed for simplicity in the derivation that
  all the observables are evaluated at the same time $t_f$, the final
  result (\ref{eq:Cn-trees}) remains valid for measurements at more
  general spacetime locations $x_1,\cdots,x_n$. The only limitation is
  that all the separations between these points should be space-like,
  $(x_i-x_j)^2<0$ if $i\not= j$, in order to avoid that the
  measurement performed at one point influences the outcome of the
  measurement performed at another point.

\item It is possible to evaluate correlation functions where different
  observables are evaluated at each point $x_i$, by having one type of
  node for each observable in the tree representation of
  eq.~(\ref{eq:Cn-trees}).

\item One can easily replace the initial vacuum state by any coherent
  state. Instead of setting the initial field $\Phi_{\rm ini}$ (and
  its first time derivative) to zero after evaluating the derivatives
  corresponding to the links in the trees of eq.~(\ref{eq:Cn-trees}),
  one would have to set them to the values of $\Phi_{\rm ini}$ and
  $\partial_0 \Phi_{\rm ini}$ that correspond to the coherent state of
  interest (see the appendix \ref{app:coherent} for more details).

\item Another extension is to consider a mixed state as initial
  state. If this state is highly occupied, the strong field
  approximation is also satisfied, as explained in the appendix
  \ref{app:mixed}.

\end{itemize}

\section{Beyond the strong field approximation}
\label{sec:beyond}
In the section \ref{sec:expansion}, we have obtained the complete tree
level result for the $1$-point (eq.~(\ref{eq:trees-1})) and $2$-point
(eq.~(\ref{eq:C12})) functions, and one readily sees that they
coincide with the result of the strong field approximation derived in
the previous section (eq.~(\ref{eq:Cn-trees}) for $n=1$ and $n=2$,
respectively). However, as we shall see now, for the $3$-point
function and beyond, the strong field approximation does not include
all the tree level contributions. This is expected
from the fact that the strong field approximation neglects some terms
in the equations of motion for the fields $\phi_1$ and $\phi_2$. In
this section, we work out the full tree level result for the $3$-point
function, in order to clarify which terms are missed by the strong
field approximation.  Firstly, let us recall here the result for
${\cal C}_{\{123\}}$ in the strong field approximation: \setbox1\hbox
to 26mm{\resizebox*{26mm}{!}{\includegraphics{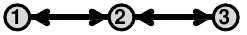}}} \setbox2\hbox
to 26mm{\resizebox*{26mm}{!}{\includegraphics{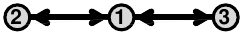}}} \setbox3\hbox
to 26mm{\resizebox*{26mm}{!}{\includegraphics{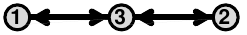}}}
\begin{equation}
  {\cal C}_{\{123\}}\Big|_{{\rm strong}\atop{\rm field}}=
  \raise -1mm\box1 + \raise -1mm\box2 + \raise -1mm\box3\; .
  \label{eq:C123-sfr}
\end{equation}
(The first of these contributions corresponds to the figure \ref{fig:LC-sfr}.)

Let us now calculate in full the $3$-point function at tree level,
including the contributions that are beyond the strong field
approximation. To that effect, we must return to the original
equations of motion (\ref{eq:eoms-phi4}), and expand them to second
order in $z$, which leads to
\begin{eqnarray}
  &&\Big[\square_x\!+\!m^2\!+\!\tfrac{g^2}{2}\big(\phi_2^{(0)}(x)\big)^2\Big]\,\phi_1^{(2)}(x;\x_1,\x_2)=0\; ,\nonumber\\
  &&\Big[\square_x\!+\!m^2\!+\!\tfrac{g^2}{2}\big(\phi_2^{(0)}(x)\big)^2\Big]\,\phi_2^{(2)}(x;\x_1,\x_2)
     =
     -g^2 \phi_2^{(0)}(x)\phi_2^{(1)}(x;\x_1)\phi_2^{(1)}(x;\x_2)\nonumber\\
  &&\qquad\qquad\qquad\qquad\qquad\qquad\qquad
     \underline{-\tfrac{g^2}{4}\phi_2^{(0)}(x)\phi_1^{(1)}(x;\x_1)\phi_1^{(1)}(x;\x_2)}\; ,
\end{eqnarray}
where we have systematically used the fact that $\phi_1^{(0)}\equiv 0$
in order to eliminate a few terms. The underlined term in the second
equation is the only one that comes from the $\phi_1^3\phi_2$
interaction term in eq.~(\ref{eq:LL-phi4}), that we had neglected in
the strong field approximation.  The boundary conditions obeyed by
these second-order coefficients at the final time read
\begin{eqnarray}
  &&  \phi_1^{(2)}(t_f,\x;\x_1,\x_2)=0\; ,\nonumber\\
  &&\partial_0 \phi_1^{(2)}(t_f,\x;\x_1,\x_2)=
     \delta(\x-\x_1)\,{\cal O}''(\phi_2^{(0)}(x_1))\,\phi_2^{(1)}(x,\x_2)\nonumber\\
  &&\qquad\qquad\qquad\qquad\quad
     + \delta(\x-\x_2)\,{\cal O}''(\phi_2^{(0)}(x_2))\,\phi_2^{(1)}(x,\x_1)\; ,
\end{eqnarray}
while those at the initial time relate their Fourier coefficients as follows
\begin{eqnarray}
  &&{\wt{\bs\phi}}_2^{(2)(+)}(\k;\x_1,\x_2)+\frac{1}{2}\,{\wt{\bs\phi}}_1^{(2)(+)}(\k;\x_1,\x_2)=0\;, \nonumber\\
  &&{\wt{\bs\phi}}_2^{(2)(-)}(\k;\x_1,\x_2)-\frac{1}{2}\,{\wt{\bs\phi}}_1^{(2)(-)}(\k;\x_1,\x_2)=0\;.
     \label{eq:bc-ti-phi22}
\end{eqnarray}

Let us recall now that the strong field approximation is exact for all
the coefficients of lesser order, i.e. $\phi_{1,2}^{(0)}$ and
$\phi_{1,2}^{(1)}$. Therefore, since the equation for $\phi_1^{(2)}$
does not contain any term coming from the vertex $\phi_2\phi_1^3$, its
solution is identical to the result of the strong field
approximation.  Let us now focus on the equation for
$\phi_2^{(2)}$. It has the structure of a linear equation of motion
with the terms in the right hand side playing the role of source
terms, since they do not contain $\phi_2^{(2)}$ itself. Therefore, we
may decompose the solution as the sum of two terms; a term that solves
the homogeneous (i.e. without source) equation and obeys the
non-trivial boundary conditions (\ref{eq:bc-ti-phi22}), and a term
that solves the full equation with a trivial null initial condition:
\begin{equation}
\phi_2^{(2)}\equiv \psi_2^{(2)}+\xi_2^{(2)}\; ,
\end{equation}
with
\begin{eqnarray}
  &&
     \Big[\square_x\!+\!m^2\!+\!\tfrac{g^2}{2}\big(\phi_2^{(0)}(x)\big)^2\Big]\,\psi_2^{(2)}=0\; ,\nonumber\\
  && {\wt{\bs\psi}}_2^{(2)(+)}+\frac{1}{2}\,{\wt{\bs\phi}}_1^{(2)(+)}=0\;, \quad
     {\wt{\bs\psi}}_2^{(2)(-)}-\frac{1}{2}\,{\wt{\bs\phi}}_1^{(2)(-)}=0 \quad \mbox{(at $t_i$)}\; ,
\end{eqnarray}
and
\begin{eqnarray}
  &&\Big[\square_x\!+\!m^2\!+\!\tfrac{g^2}{2}\big(\phi_2^{(0)}(x)\big)^2\Big]\,\xi_2^{(2)}
     =
     -g^2 \phi_2^{(0)}\phi_2^{(1)}\phi_2^{(1)}
     \underline{-\tfrac{g^2}{4}\phi_2^{(0)}\phi_1^{(1)}\phi_1^{(1)}}\; ,
     \nonumber\\
  &&\xi_2^{(2)}=\partial_0 \xi_2^{(2)}=0\quad \mbox{(at $t_i$)}\; .
\end{eqnarray}
Concerning $\psi_2^{(2)}$, the equation of motion does not contain any
term coming from the $\phi_2\phi_1^3$ vertex, and all the objects that
appear in the equation of motion and boundary conditions are exact in
the strong field approximation. Therefore, $\psi_2^{(2)}$ is itself
correctly given by this approximation. For $\xi_2^{(2)}$, we can write
the following solution:
\begin{eqnarray}
  &&\xi_2^{(2)}(x;\x_1,\x_2)
  =
  -g^2\int \rmd^4y\;G_{_R}(x,y)\, \phi_2^{(0)}(y)
     \Big[\phi_2^{(1)}(y;\x_1)\phi_2^{(1)}(y;\x_2)\nonumber\\
  &&\qquad\qquad\qquad\qquad\qquad\qquad\qquad
  +\underline{\tfrac{1}{4}\phi_1^{(1)}(y;\x_1)\phi_1^{(1)}(y;\x_2)}
  \Big]\; ,
\end{eqnarray}
where $G_{_R}(x,y)$ is the retarded propagator dressed by the
background field $\phi_2^{(0)}$. The first term is already contained in
the strong field approximation since it does not involve the
$\phi_2\phi_1^3$ vertex, but the second term is present only 
beyond this approximation. Therefore, the only tree level term in the
$3$-point correlation function that is not contained in the strong
field approximation is
\begin{eqnarray}
  {\cal C}_{\{123\}}\Big|_{{\rm beyond}\atop{\rm SFA}}&=&
                                                          -\frac{ig^2}{4}\;{\cal O}'(\phi_2^{(0)}(x_1)){\cal O}'(\phi_2^{(0)}(x_2)){\cal O}'(\phi_2^{(0)}(x_3))\nonumber\\
                                                      &&\qquad\times\int \rmd^4y\;G_{_R}(x_3,y)\, \phi_2^{(0)}(y)\phi_1^{(1)}(y;\x_1)\phi_1^{(1)}(y;\x_2)\nonumber\\
  &=&\frac{ig^2}{4}\;{\cal O}'(\phi_2^{(0)}(x_1)){\cal O}'(\phi_2^{(0)}(x_2)){\cal O}'(\phi_2^{(0)}(x_3))\nonumber\\
                                                      &&\qquad\times\int \rmd^4y\;G_{_R}(x_1,y)G_{_R}(x_2,y)G_{_R}(x_3,y)\, \phi_2^{(0)}(y)\; ,
\end{eqnarray}
where we have used the explicit form (\ref{eq:phi11}) of
$\phi_1^{(1)}$ in order to rewrite it in a completely symmetric form
in the second equality. The peculiarity of this contribution is that
the correlation is created in the bulk by the self-interactions of the
fields, instead of the pairwise initial state correlations that we
have encountered in the strong field approximation. Generalizing the
diagrammatic representation of eq.~(\ref{eq:Cn-trees}), such a term
 may be represented as follows:\setbox1\hbox to
13.5mm{\resizebox*{13.5mm}{!}{\includegraphics{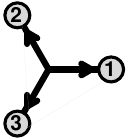}}}
\begin{equation}
  {\cal C}_{\{123\}}\Big|_{{\rm beyond}\atop{\rm SFA}}=
  \;\raise -7mm\box1\; .
  \label{eq:C123-bsfr}
\end{equation}
(Note that in this representation, the power of the classical field
$\phi_2^{(0)}(y)$ that accompanies the vertex does not appear
explicitly.)  The causal structure of this term is illustrated in the
figure \ref{fig:LC-bsfr}:
\begin{figure}[htbp]
  \begin{center}
    \resizebox*{11cm}{!}{\includegraphics{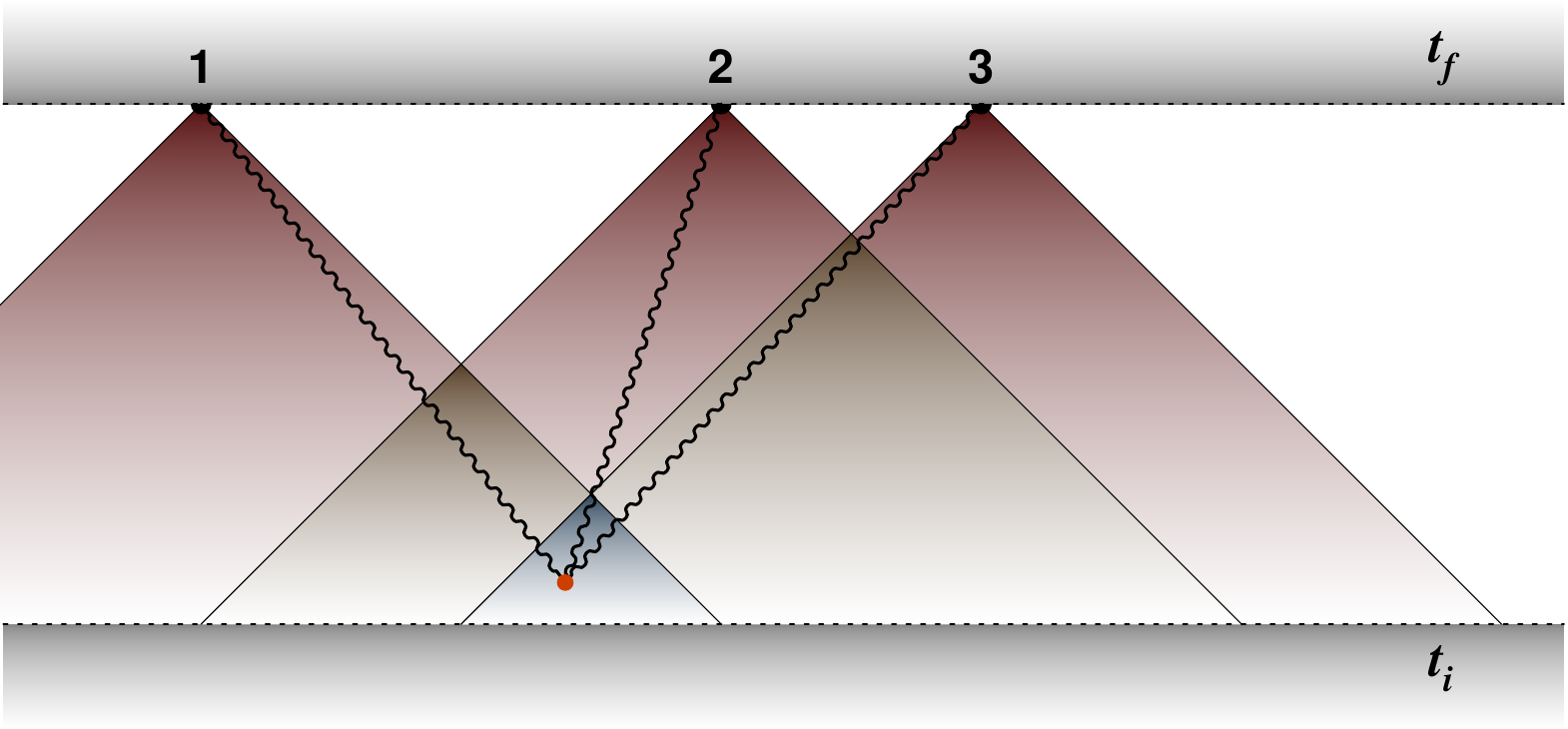}}
  \end{center}
  \caption{\label{fig:LC-bsfr} Causal structure of the extra
    contribution to the $3$-point correlation function that arises
    beyond the strong field approximation. In this contribution, the
    correlation is produced in the bulk, from the vertex
    $\phi_1^3\phi_2$. Causality implies that this vertex must be
    located in the intersection of the three light-cones. The red dot
    contains a power of the classical field $\phi_2^{(0)}(y)$.}
\end{figure}
because of the three retarded propagators, the vertex $y$ must be
located inside the overlap of the three past light-cones of summits
$x_{1,2,3}$, but it is not tied to the initial surface.

To close this section, let us compare the orders of magnitude of the
contributions of eq.~(\ref{eq:C123-sfr}) and of
eq.~(\ref{eq:C123-bsfr}) in the presence of the instabilities
discussed in the section \ref{sec:insta}. On the one hand, we have
\setbox1\hbox
to 26mm{\resizebox*{26mm}{!}{\includegraphics{C123-1}}}
\begin{equation}
    \raise -1mm\box1 \sim\; g^4 \; e^{4\mu(t_f-t_i)}\; ,
  \label{eq:C123-sfr1}
\end{equation}
where the factor $g^4$ is relative to the completely disconnected
$3$-point function (each derivative $\delta/\delta\Phi_{\rm ini}$
brings a factor $g$ when the background field $\Phi$ is of order
$g^{-1}$). Regarding the contribution of eq.~(\ref{eq:C123-bsfr}),
each link brings a factor $g$ since it replaces a field $\Phi$, and
the factor $g^2\,\phi_2^{(0)}$ is of order $g$. Regarding the time
dependence, each retarded propagator behaves as
\begin{equation}
G_{_R}(x_1,y)\sim e^{\mu(t_f-y^0)}\; .
\end{equation}
Therefore, we have\setbox1\hbox to
13.5mm{\resizebox*{13.5mm}{!}{\includegraphics{C123-bsfr}}}
\begin{equation}
  \raise -7mm\box1\sim \; g^4\;e^{3\mu(t_f-y^0)}\lesssim g^4\;e^{3\mu(t_f-t_i)}\; .
  \label{eq:C123-bsfr1}
\end{equation}
We see that the terms beyond the strong field approximation have the
same order in $g$, but are exponentially suppressed by a factor of
order $\exp(-\mu(t_f-t_i))$.

\section{Summary and conclusions}
\label{sec:concl}
In this paper, we have studied the correlation function between an
arbitrary number of observables measured at equal times (or more
generally at points with space-like separations) in quantum field
theory. Our main focus has been the strong field regime, that arises
for instance when the fields are driven by a large external source and
their classical equations of motion are subject to instabilities.

Firstly, we have constructed a generating functional that encapsulates
all these correlation functions by coupling the observables to a
fictitious source $z(\x)$. Using the retarded-advanced basis of the
in-in formalism, we have shown that it can be expressed at tree level
in terms of a pair of fields that obey coupled equations of motion and
non-trivial boundary conditions that depend on the observables. At the
first two orders in the fictitious source (i.e. for the $1$-point and
$2$-point correlation functions), the results can be expressed in terms
of the retarded classical solution of the field equation of motion,
and its functional derivative with respect to its initial condition.

For these two lowest orders, the result we have obtained is in fact
exact at tree level, and does not rely on having strong fields. Beyond
these low orders, this direct approach is very cumbersome to extend
systematically. In order to circumvent this difficulty, we have
introduced the strong field approximation, thanks to which one can
formally solve the equations of motion with the appropriate boundary
conditions, in the form of an implicit functional equation for the
first derivative of the generating functional. From this functional
relation, we have found that all the correlation functions are
expressible in terms of the classical field and its derivatives with
respect to the initial condition, generalizing the results for the
$1$-point and $2$-point functions. Moreover, the expressions can be
systematically represented by trees, where the nodes are the
observables and the links are pairs of derivatives with respect to the
initial condition of the classical field. Physically, these links
correspond to correlations induced by fluctuations in the initial
state.

In the last section, we departed from the strong field approximation
and considered the tree-level $3$-point correlation function in
full. We found an additional tree-level term that does not exist in
the strong field regime, corresponding to correlations created in
the bulk by the interactions themselves. We expect that such terms
(and more complicated ones) exist in all $n$-point tree-level
correlation functions for any $n\ge 3$, beyond the strong field
regime.

\section*{Acknowledgements}
I would like to thank F. Vernizzi and P. Creminelli for discussions on
cosmological perturbations, and E. Guitter for very useful
explanations on the combinatorics of trees. This work is support by
the Agence Nationale de la Recherche project ANR-16-CE31-0019-01.

\appendix

\section{Some properties of the mode functions}
\label{app:modes}
In the section \ref{sec:modfunc}, we have introduced a basis of
solutions for the linear space of solutions for a partial differential
equation of the form
\begin{equation}
  \Big[\square_x+m^2-{\cal L}''_{\rm int}(\varphi(x))\Big]\,a(x)=0\; ,
  \label{eq:leom}
\end{equation}
where $\varphi(x)$ is some real background field.  This basis was
defined as a set of solutions $\{a_{\pm\k}(x)\}$ labeled by a momentum
$\k$, whose initial condition is a plane wave of positive (in the case
of $a_{+\k}$) or negative (in the case of $a_{-\k}$) frequency:
\begin{equation}
a_{\pm\k}(x)\empile{\to}\over{x^0\to t_i}e^{\mp i k\cdot x}\; .
\end{equation}
Any solution of the equation (\ref{eq:leom}) is a linear
superposition of the functions $a_{\pm\k}$.

Given a solution $a(x)$, let us define the following two components vectors
\begin{equation}
  \big|{\bs a}\big)\equiv \begin{pmatrix} a\\ \dot{a}\end{pmatrix}\; ,
  \qquad
  \big({\bs a}\big|\equiv i\begin{pmatrix} -\dot{a}^* & a^*\end{pmatrix}\; .
\end{equation}
Then, from two solutions $a_1$ and $a_2$ of eq.~(\ref{eq:leom}), we may
define the following inner product
\begin{equation}
  \big({\bs a}_1\big|{\bs a}_2\big)\equiv i\int d^3\x\;\Big[
  a_1^*(x)\,\dot{a}_2(x)-\dot{a}_1^*(x)\,a_2(x)
  \Big]\; ,
\end{equation}
reminiscent of the Wronskian for solutions of ordinary differential
equations. This product is Hermitian,
\begin{equation}
\big({\bs a}_2\big|{\bs a}_1\big)=\big({\bs a}_1\big|{\bs a}_2\big)^*\; ,
\end{equation}
and constant in time:
\begin{equation}
\partial_0\big({\bs a}_1\big|{\bs a}_2\big) =0\; .
\end{equation}
Its value is therefore completely determined by the initial conditions
for the solutions $a_1$ and $a_2$. In the case of the solutions
$a_{\pm\k}$ introduced above, an explicit calculation gives
\begin{eqnarray}
  &&\big({\bs a}_{+\k}\big|{\bs a}_{+\k'}\big)=(2\pi)^3\,2E_\k\,\delta(\k-\k')\; ,
     \nonumber\\
  &&\big({\bs a}_{-\k}\big|{\bs a}_{-\k'}\big)=-(2\pi)^3\,2E_\k\,\delta(\k-\k')\; ,
     \nonumber\\
  &&\big({\bs a}_{+\k}\big|{\bs a}_{-\k'}\big)=0\; .
\end{eqnarray}

A generic solution $a$ of eq.~(\ref{eq:leom}) can be decomposed as follows
\begin{equation}
  \big|{\bs a}\big)=\int \frac{\rmd^3\k}{(2\pi)^3 2E_\k}\;\Big[
  \gamma_{+\k}\;\big|{\bs a}_{+\k}\big)+\gamma_{-\k}\;\big|{\bs a}_{-\k}\big)
  \Big]\; ,
  \label{eq:decomp}
\end{equation}
and from the inner products between the mode functions we readily see
that the coefficients of this linear decomposition are given by
\begin{equation}
  \gamma_{+\k} = \big({\bs a}_{+\k}\big|{\bs a}\big)\; ,\quad
  \gamma_{-\k} = -\big({\bs a}_{-\k}\big|{\bs a}\big)\; .
\end{equation}
Note that in the decomposition (\ref{eq:decomp}), the coefficients
$\gamma_{\pm\k}$ are constant, and the time dependence is carried by
the mode functions $a_{\pm\k}$. Therefore, we can write
\begin{equation}
  \big|{\bs a}\big)=\int \frac{\rmd^3\k}{(2\pi)^3 2E_\k}\;\Big[
  \big|{\bs a}_{+\k}\big)\big({\bs a}_{+\k}\big|{\bs a}\big)
  -
  \big|{\bs a}_{-\k}\big)\big({\bs a}_{-\k}\big|{\bs a}\big)
  \Big]\; .
\end{equation}
Since this is true for any solution $a$, the following relationship
must in fact be true
\begin{equation}
  \int \frac{\rmd^3\k}{(2\pi)^3 2E_\k}\;\Big[
  \big|{\bs a}_{+\k}\big)\big({\bs a}_{+\k}\big|
  -
  \big|{\bs a}_{-\k}\big)\big({\bs a}_{-\k}\big|
  \Big]=\begin{pmatrix}1&0\\0&1\\\end{pmatrix}\; .
\end{equation}
Reinstating coordinates, this identity reads
\begin{eqnarray}
  &&
     \int_k 
  \begin{pmatrix}
    a_{+\k}(x)\dot{a}_{-\k}(y)-a_{-\k}(x)\dot{a}_{+\k}(y)&
    a_{-\k}(x)a_{+\k}(y)-a_{+\k}(x)a_{-\k}(y)\\
    \dot{a}_{+\k}(x)\dot{a}_{-\k}(y)-\dot{a}_{-\k}(x)\dot{a}_{+\k}(y)&
    \dot{a}_{-\k}(x)a_{+\k}(y)-\dot{a}_{+\k}(x)a_{-\k}(y)\\
  \end{pmatrix}\nonumber\\
  &&\qquad\qquad\qquad\qquad\qquad
     \empile{=}\over{x^0=y^0}i\,\delta(\x-\y)\begin{pmatrix}1&0\\0&1\\\end{pmatrix}\; .
  \label{eq:modes-ret}
\end{eqnarray}
These identities are justification for eqs.~(\ref{eq:phi11}) and (\ref{eq:phi1}).

\section{In-in formalism for an initial coherent state}
\label{app:coherent}
A coherent state can be defined from the perturbative in-vacuum as
follows
\begin{equation}
  \big|\chi\big>\equiv {\cal N}_\chi \;
  \exp\Big\{\int_\k \chi(\k)\,a_{\rm in}^\dagger(\k)\Big\}\;\big|0{}_{\rm in}\big>\; ,
\end{equation}
where $\chi(\k)$ is a function of 3-momentum and ${\cal N}_\chi$ a
normalization constant adjusted so that
$\big<\chi\big|\chi\big>=1$. From the canonical commutation relation
\begin{equation}
\big[a_{\rm in}(\p),a_{\rm in}^\dagger(\q)\big]=(2\pi)^3 \,2E_\p\,\delta(\p-\q)\; ,
\end{equation}
it is easy to check the following 
\begin{eqnarray}
  && a_{\rm in}(\p)\, \big|\chi\big>=\chi(\p)\,\big|\chi\big>\; ,
     \nonumber\\
  &&
\big|{\cal N}_\chi\big|^2 = \exp\Big\{-\int_\k \big|\chi(\k)\big|^2\Big\}\; .
\end{eqnarray}
The first equation tells us that $\big|\chi\big>$ is an eigenstate of
annihilation operators, which is another definition of coherent
states, and the second one provides the value of the normalization
constant.

Consider now the generating functional for the in-in formalism in this
coherent state,
\begin{eqnarray}
  Z_\chi[\eta]&\equiv&\big<\chi\big|
  {\rm P}\,\exp i\int_{\cal C}\rmd^4x\;
  \eta(x)\phi(x)
  \big|\chi\big>
                       \nonumber\\
  &=&\big<\chi\big|
  {\rm P}\,\exp i\int_{\cal C}\rmd^4x\;\Big[
  {\cal L}_{\rm  int}(\phi_{\rm in}(x))+\eta(x)\phi_{\rm in}(x)\Big]
      \big|\chi\big>\; ,
      \label{eq:Zchi-0}
\end{eqnarray}
where $\eta(x)$ is a fictitious source that lives on the closed-time
contour ${\cal C}$ introduced after eq.~(\ref{eq:Fin}). The first step
is to factor out the interactions as follows:
\begin{equation}
  Z_\chi[\eta]=
  \exp i\int_{\cal C} \rmd^4x\; {\cal L}_{\rm int}\Big(\frac{\delta}{i\delta \eta(x)}\Big)
  \;\underbrace{\big<\chi\big|
  {\rm P}\,\exp i\int_{\cal C}\rmd^4x\;
  \eta(x)\phi_{\rm in}(x)
  \big|\chi\big>}_{Z_{\chi 0}[\eta]}\; .
\end{equation}
A first use of the Baker-Campbell-Hausdorff formula
enables one to remove the path ordering, giving
\begin{eqnarray}
&&Z_{\chi 0}[\eta]
  =
  \big<\chi\big|
  \exp i\int_{\cal C}\rmd^4x\;
  \eta(x)\phi_{\rm in}(x)
   \big|\chi\big>\nonumber\\
  &&\quad\times
  \exp\Big\{ -\frac{1}{2}\int_{\cal C}\rmd^4x \rmd^4y\;\eta(x)\eta(y)\;\theta_c(x^0-y^0)\;
  \big[\phi_{\rm in}(x),\phi_{\rm in}(y)\big]\Big\}\; ,
\end{eqnarray}
where $\theta_c(x^0-y^0)$ generalizes the step function to the ordered
contour ${\cal C}$. Note that the factor on the second line is a
commuting number and thus can be removed from the expectation value.  A
second application of the Baker-Campbell-Hausdorff formula
allows to normal-order the first factor. Decomposing the in-field as
follows,
\begin{equation}
  \phi_{\rm in}(x)\equiv
  \underbrace{\int_\k a_{\rm in}(\k)\,e^{-ik\cdot x}}_{\phi_{\rm in}^{(-)}(x)}
  +
  \underbrace{\int_\k a_{\rm in}^\dagger(\k)\,e^{+ik\cdot x}}_{\phi_{\rm in}^{(+)}(x)}\; ,
\end{equation}
we obtain
\begin{eqnarray}
&&Z_{\chi 0}[\eta]
   =
   \big<\chi\big|\exp \Big\{i\int_{\cal C}\rmd^4x \;\eta(x)\phi_{\rm in}^{(+)}(x)\Big\}
   \exp \Big\{i\int_{\cal C}\rmd^4y \;\eta(y)\phi_{\rm in}^{(-)}(y)\Big\}\big|\chi\big>
   \nonumber\\
  &&\quad\times
   \exp\Big\{+\frac{1}{2}\int_{\cal C}\rmd^4x \rmd^4y\;\eta(x)\eta(y)\,\big[\phi_{\rm in}^{(+)}(x),\phi_{\rm in}^{(-)}(y)\big]\Big\}
   \nonumber\\
  &&\quad\times
  \exp\Big\{ -\frac{1}{2}\int_{\cal C}\rmd^4x \rmd^4y\;\eta(x)\eta(y)\;\theta_c(x^0-y^0)\;
     \big[\phi_{\rm in}(x),\phi_{\rm in}(y)\big]\Big\}\; .
     \label{eq:tmp1}
\end{eqnarray}
The factor of the first line can be evaluated by using the fact that
the coherent state is an eigenstate of annihilation operators:
\begin{eqnarray}
  &&
\big<\chi\big|\exp \Big\{i\int_{\cal C}\rmd^4x \;\eta(x)\phi_{\rm in}^{(+)}(x)\Big\}
     \exp \Big\{i\int_{\cal C}\rmd^4y \;\eta(y)\phi_{\rm in}^{(-)}(y)\Big\}\big|\chi\big>
     \nonumber\\
  &&\quad
  =\exp\Big\{ i\int_{\cal C}\rmd^4x\;\eta(x)\;\underbrace{\int_\k\Big(\chi(\k)e^{-ik\cdot x}+\chi^*(\k)e^{+ik\cdot x}\Big)}_{\Phi_\chi(x)}\Big\}\; .
\end{eqnarray}
We denote $\Phi_\chi(x)$ the field obtained by substituting the
creation and annihilation operators of the in-field by $\chi^*(\k)$
and $\chi(\k)$ respectively. Note that this is no longer an operator,
but a (real valued) commuting field. Moreover, because it is a linear
superposition of plane waves, this field is a free field:
\begin{equation}
(\square_x+m^2)\,\Phi_\chi(x)=0\; .
\end{equation}
The second and third factors of eq.~(\ref{eq:tmp1}) are commuting
numbers, provided we do not attempt to disassemble the
commutators. Using the decomposition of the in-field in terms of
creation and annihilation operators, and the canonical commutation
relation of the latter, we obtain
\begin{eqnarray}
  &&
  \theta_c(x^0-y^0)\;
  \big[\phi_{\rm in}(x),\phi_{\rm in}(y)\big]
  -
  \big[\phi_{\rm in}^{(+)}(x),\phi_{\rm in}^{(-)}(y)\big]
     \nonumber\\
  &&\qquad
  =
  \underbrace{\theta_c(x^0-y^0)\int_\k e^{-ik\cdot(x-y)}
  +
  \theta_c(y^0-x^0)\int_\k e^{+ik\cdot(x-y)}}_{G_c^0(x,y)}\; ,
\end{eqnarray}
which is nothing but the usual bare path-ordered propagator
$G_c^0(x,y)$. Collecting all the factors, the generating functional
for path-ordered Green's functions in the in-in formalism with an
initial coherent state reads
\begin{eqnarray}
  Z_\chi[\eta]&=&
  \exp\Big\{ i\int_{\cal C} \rmd^4x\; {\cal L}_{\rm int}\Big(\frac{\delta}{i\delta \eta(x)}\Big)\Big\}\;
  \underline{\exp\Big\{ i\int_{\cal C}\rmd^4x\;\eta(x)\,\Phi_\chi(x)\Big\}}
  \nonumber\\
  &&\qquad\times
     \exp\Big\{-\frac{1}{2}\int_{\cal C}\rmd^4x \rmd^4y\;\eta(x)\eta(y)\;G_c^0(x,y)\Big\}\; .
     \label{eq:Zchi-1}
\end{eqnarray}
We see that it differs from the corresponding functional with the
perturbative vacuum\footnote{The vacuum initial state corresponds to
  the function $\chi(\k)\equiv 0$, i.e. to $\Phi_\chi(x)$.} as initial
state only by the second factor, that we have underlined. This generating
functional is also equal to\footnote{In this transformation, we use
  the functional analogue of \begin{equation*}
    F(\partial_x)\,e^{\alpha x}\, G(x) = e^{\alpha x}\,F(\alpha+\partial_x)\,G(x)\;
    .
  \end{equation*}}
\begin{eqnarray}
  Z_\chi[\eta]&=&\exp\Big\{ i\int_{\cal C}\rmd^4x\;\eta(x)\,\Phi_\chi(x)\Big\}\;
  \exp\Big\{ i\int_{\cal C} \rmd^4x\; {\cal L}_{\rm int}\Big(\Phi_\chi(x)+\frac{\delta}{i\delta \eta(x)}\Big)\Big\}\;
  \nonumber\\
  &&\qquad\times
     \exp\Big\{-\frac{1}{2}\int_{\cal C}\rmd^4x \rmd^4y\;\eta(x)\eta(y)\;G_c^0(x,y)\Big\}\; .
     \label{eq:Zchi-2}
\end{eqnarray}
The first factor amounts to shifting the fields by $\Phi_\chi(x)$. The
simplest way to see this is to write
\begin{equation}
\phi\equiv \Phi_\chi+\zeta\; .
\end{equation}
In the definition (\ref{eq:Zchi-0}), this leads to
\begin{equation}
  Z_\chi[\eta]=\exp\Big\{i\int_{\cal C}\rmd^4\;\eta(x)\,\Phi_\chi(x)\Big\}
  \;\big<\chi\big|
  {\rm P}\,\exp i\int_{\cal C}\rmd^4x\;
  \eta(x)\,\zeta(x)
  \big|\chi\big>\; ,
\end{equation}
where the second factor in the right hand side is the generating
functional for correlators of $\zeta$. Comparing with
eq.~(\ref{eq:Zchi-2}), we see that the generating functional for
$\zeta$ is identical to the vacuum one, except that the argument
$\phi$ of the interaction Lagrangian is replaced by $\Phi_\chi+\zeta$:
\begin{equation}
  {\cal L}_{\rm int}(\phi)\quad\to\quad
  {\cal L}_{\rm int}(\Phi_\chi+\zeta)\; .
\end{equation}
In other words, the field $\zeta$ appears to be coupled to an external
source and to a background field. Note that in the in-in formalism,
the change $\phi\to\Phi_\chi+\zeta$ applies equally to the fields
$\phi_\pm$ on both branches of the time contour. Therefore, for the
fields $\phi_{1,2}$ in the retarded-advanced basis, we have
\begin{equation}
  \phi_1=\zeta_1\quad,\qquad
  \phi_2=\Phi_\chi+\zeta_2\; ,
\end{equation}
and the integral equations (\ref{eq:int}) that determine these fields
at tree level become
\begin{eqnarray}
\zeta_1(x)
&=&
i\int_\Omega \rmd^4y\;
    G_{12}^0(x,y)\,\frac{\partial{\bs L}_{\rm int}(\zeta_1,\Phi_\chi+\zeta_2)}{\partial \zeta_2(y)}
\nonumber\\
&&\qquad
+\int_{t_f} \rmd^3\y\;
G_{12}^0(x,y)\;z(\y)\;{\cal O}'(\Phi_\chi(y)+\zeta_2(y))\; ,\nonumber\\
\zeta_2(x)
&=&
i\int_\Omega \rmd^4y\;\Big\{
G_{21}^0(x,y)\frac{\partial{\bs L}_{\rm int}(\zeta_1,\Phi_\chi\!+\!\zeta_2)}{\partial \zeta_1(y)}
\!+\!
G_{22}^0(x,y)\frac{\partial{\bs L}_{\rm int}(\zeta_1,\Phi_\chi\!+\!\zeta_2)}{\partial \zeta_2(y)}
\Big\}\nonumber\\
&&\qquad
+\int_{t_f} \rmd^3\y\;
G_{22}^0(x,y)\;z(\y)\;{\cal O}'(\Phi_\chi(y)+\zeta_2(y))\; .
\label{eq:int-coh}
\end{eqnarray}
Using $\phi_2=\Phi_\chi+\zeta_2$ and the fact that $\Phi_\chi$ is a
free field, we see that the equations of motion corresponding to these
integral equations are the same as eqs.~(\ref{eq:eoms}). The boundary
conditions at the final time read:
\begin{equation}
  \zeta_1(t_f,\x)=0\quad,\qquad
  \partial_0\zeta_1(t_t,\x)=i \,z(\x)\,{\cal O}'\big(\Phi_\chi(t_f,\x)+\zeta_2(t_f,\x)\big)\; ,
\end{equation}
while at the initial time we have the following relationship
\begin{equation}
  {\wt{\bs\zeta}}_2^{(+)}(\k)
  =-\frac{1}{2}\,{\wt{\bs\zeta}}_1^{(+)}(\k)\;,\quad
  {\wt{\bs\zeta}}_2^{(-)}(\k)
  =\frac{1}{2}\,{\wt{\bs\zeta}}_1^{(-)}(\k)
\end{equation}
among the Fourier coefficients. The derivation of the main result
(\ref{eq:master-D}) for a vacuum initial state can be reproduced
almost identically for a coherent initial state, leading to
\begin{equation}
  {\cal D}[\x_1;z]
  =\exp\Bigg\{
  \int \rmd^3\u\;z(\u)\,{\cal D}[\u;z]\;\otimes
  \Bigg\}\;{\cal O}\big(\Phi(t_f,\x_1)\big)\Bigg|_{\Phi_{\rm ini}\equiv \Phi_\chi}\; ,
  \label{eq:master-D-coh}
\end{equation}
the only difference being that the initial value of the classical field
$\Phi$ is set to $\Phi_\chi$ instead of zero after the derivatives
with respect to $\Phi_{\rm ini}$ have been performed.

\section{In-in formalism for a Gaussian mixed state}
\label{app:mixed}
Let us consider in this section a mixed initial state described by the
following density matrix
\begin{equation}
\rho\equiv \exp \Big\{-\int_\k \beta_\k E_\k\,a^\dagger_{\rm in}(\k)a_{\rm in}(\k)\Big\}\; .
\end{equation}
In this definition, $\beta_\k$ has the same function as an inverse
temperature (by allowing it to be momentum dependent we can also
consider out-of-equilibrium systems). The expectation value in the
pure initial state of eq.~(\ref{eq:F-def}) is replaced by a trace
\begin{equation}
  \big<0{}_{\rm in}\big|\cdots \big|0{}_{\rm in}\big>
  \quad\to\quad
  \frac{{\rm Tr}\,\big(\rho\cdots\big)}{{\rm Tr}\,\big(\rho\big)}\; .
\end{equation}
How to handle this type of initial state is well known from quantum
field theory at finite temperature. The perturbative rules are
identical to those exposed in the figure \ref{fig:rules}, but the
propagators of eqs.~(\ref{eq:props-vac}) should be replaced by

\begin{eqnarray}
  &&
  G_{-+}^0(x,y)=\int_\k \Big((1+f_\k)\,e^{-ik\cdot(x-y)}+f_\k\,e^{ik\cdot(x-y)}\Big)
     \;,\nonumber\\
  &&
  G_{+-}^0(x,y)=\int_\k \Big(f_\k\,e^{-ik\cdot(x-y)}+(1+f_\k)\,e^{ik\cdot(x-y)}\Big)
     \; ,\nonumber\\
  && G_{++}^0(x,y)=\theta(x^0-y^0)\,G_{-+}^0(x,y)+\theta(y^0-x^0)\,G_{+-}^0(x,y)\; ,
     \nonumber\\
  && G_{--}^0(x,y)=\theta(x^0-y^0)\,G_{+-}^0(x,y)+\theta(y^0-x^0)\,G_{-+}^0(x,y)\; ,
     \label{eq:props-mixed}
\end{eqnarray}
where we have defined
\begin{equation}
f_\k\equiv \frac{1}{e^{\beta_\k E_\k}-1}\; .
\end{equation}
In the retarded-advanced basis, the propagators $G_{12}^0$ and
$G_{21}^0$ are unmodified, but the propagator $G_{22}^0$ becomes
$f_\k$-dependent. The integral equations (\ref{eq:int}) are unaltered
(but they acquire a hidden dependence on $f_\k$ through $G_{22}^0$),
and consequently the equations of motion (\ref{eq:eoms}) are
unchanged. The boundary conditions (\ref{eq:bc-final}) at the final
time remain the same, but those at the initial time
(\ref{eq:bc-ti-ii}) are modified into
\begin{equation}
  {\wt{\bs\phi}}_{2}^{(+)}(\k)=-\Big(\frac{1}{2}+f_\k\Big)\,{\wt{\bs\phi}}_{1}^{(+)}(\k)\;,\quad
  {\wt{\bs\phi}}_{2}^{(-)}(\k)=\Big(\frac{1}{2}+f_\k\Big)\,{\wt{\bs\phi}}_{1}^{(-)}(\k)\; .
  \label{eq:bc-ti-ii-mixed}
\end{equation}
In other words, the factor $\tfrac{1}{2}$ of eqs.~(\ref{eq:bc-ti-ii}),
that can be interpreted as the zero point occupation of the vacuum,
is now replaced by the total occupation number $\tfrac{1}{2}+f_\k$ of
the mixed initial state under consideration. We see from
eqs.~(\ref{eq:bc-ti-ii-mixed}) that a large occupation number 
enhances $\phi_2$ with respect to $\phi_1$. This is another situation
where the strong field approximation introduced in the section
\ref{sec:strong} is applicable.


\providecommand{\href}[2]{#2}\begingroup\raggedright\begin{thebibliography}{10}

\bibitem{Mukhanov:1990me}
V.~F. Mukhanov, H.~A. Feldman, and R.~H. Brandenberger, ``{Theory of
  cosmological perturbations. Part 1. Classical perturbations. Part 2. Quantum
  theory of perturbations. Part 3. Extensions},''
\href{http://dx.doi.org/10.1016/0370-1573(92)90044-Z}{{\em Phys. Rept.}
  {\bfseries 215} (1992) 203--333}.

\bibitem{Weinberg:2005vy}
S.~Weinberg, ``{Quantum contributions to cosmological correlations},''
  \href{http://dx.doi.org/10.1103/PhysRevD.72.043514}{{\em Phys. Rev.}
  {\bfseries D72} (2005) 043514},
\href{http://arxiv.org/abs/hep-th/0506236}{{\ttfamily arXiv:hep-th/0506236
  [hep-th]}}.

\bibitem{Gelis:2010nm}
F.~Gelis, E.~Iancu, J.~Jalilian-Marian, and R.~Venugopalan, ``{The Color Glass
  Condensate},''
  \href{http://dx.doi.org/10.1146/annurev.nucl.010909.083629}{{\em Ann. Rev.
  Nucl. Part. Sci.} {\bfseries 60} (2010) 463--489},
\href{http://arxiv.org/abs/1002.0333}{{\ttfamily arXiv:1002.0333 [hep-ph]}}.

\bibitem{Gelis:2012ri}
F.~Gelis, ``{Color Glass Condensate and Glasma},''
  \href{http://dx.doi.org/10.1142/S0217751X13300019}{{\em Int. J. Mod. Phys.}
  {\bfseries A28} (2013) 1330001},
\href{http://arxiv.org/abs/1211.3327}{{\ttfamily arXiv:1211.3327 [hep-ph]}}.

\bibitem{Schwinger1961}
J.~Schwinger, ``{Brownian Motion of a Quantum Oscillator},''
  \href{http://dx.doi.org/10.1063/1.1703727}{{\em J. Math. Phys.} {\bfseries 2}
  (1961) 407}.

\bibitem{Bakshi:1962dv}
P.~M. Bakshi and K.~T. Mahanthappa, ``{Expectation value formalism in quantum
  field theory. 1.},''
\href{http://dx.doi.org/10.1063/1.1703883}{{\em J. Math. Phys.} {\bfseries 4}
  (1963) 1--11}.

\bibitem{Bakshi:1963bn}
P.~M. Bakshi and K.~T. Mahanthappa, ``{Expectation value formalism in quantum
  field theory. 2.},''
\href{http://dx.doi.org/10.1063/1.1703879}{{\em J. Math. Phys.} {\bfseries 4}
  (1963) 12--16}.

\bibitem{Keldysh:1964ud}
L.~V. Keldysh, ``{Diagram technique for nonequilibrium processes},'' {\em Zh.
  Eksp. Teor. Fiz.} {\bfseries 47} (1964) 1515--1527.
[Sov. Phys. JETP20,1018(1965)].

\bibitem{Chou:1984es}
K.-c. Chou, Z.-b. Su, B.-l. Hao, and L.~Yu, ``{Equilibrium and Nonequilibrium
  Formalisms Made Unified},''
\href{http://dx.doi.org/10.1016/0370-1573(85)90136-X}{{\em Phys. Rept.}
  {\bfseries 118} (1985) 1}.

\bibitem{Jordan:1986ug}
R.~D. Jordan, ``{Effective Field Equations for Expectation Values},''
\href{http://dx.doi.org/10.1103/PhysRevD.33.444}{{\em Phys. Rev.} {\bfseries
  D33} (1986) 444--454}.

\bibitem{Gelis:2006yv}
F.~Gelis and R.~Venugopalan, ``{Particle production in field theories coupled
  to strong external sources},''
  \href{http://dx.doi.org/10.1016/j.nuclphysa.2006.07.020}{{\em Nucl. Phys.}
  {\bfseries A776} (2006) 135--171},
\href{http://arxiv.org/abs/hep-ph/0601209}{{\ttfamily arXiv:hep-ph/0601209
  [hep-ph]}}.

\bibitem{Gelis:2006cr}
F.~Gelis and R.~Venugopalan, ``{Particle production in field theories coupled
  to strong external sources. II. Generating functions},''
  \href{http://dx.doi.org/10.1016/j.nuclphysa.2006.08.015}{{\em Nucl. Phys.}
  {\bfseries A779} (2006) 177--196},
\href{http://arxiv.org/abs/hep-ph/0605246}{{\ttfamily arXiv:hep-ph/0605246
  [hep-ph]}}.

\bibitem{Gelis:2008rw}
F.~Gelis, T.~Lappi, and R.~Venugopalan, ``{High energy factorization in
  nucleus-nucleus collisions},''
  \href{http://dx.doi.org/10.1103/PhysRevD.78.054019}{{\em Phys. Rev.}
  {\bfseries D78} (2008) 054019},
\href{http://arxiv.org/abs/0804.2630}{{\ttfamily arXiv:0804.2630 [hep-ph]}}.

\bibitem{Gelis:2008ad}
F.~Gelis, T.~Lappi, and R.~Venugopalan, ``{High energy factorization in
  nucleus-nucleus collisions. II. Multigluon correlations},''
  \href{http://dx.doi.org/10.1103/PhysRevD.78.054020}{{\em Phys. Rev.}
  {\bfseries D78} (2008) 054020},
\href{http://arxiv.org/abs/0807.1306}{{\ttfamily arXiv:0807.1306 [hep-ph]}}.

\bibitem{Weinberg:2008mc}
S.~Weinberg, ``{A Tree Theorem for Inflation},''
  \href{http://dx.doi.org/10.1103/PhysRevD.78.063534}{{\em Phys. Rev.}
  {\bfseries D78} (2008) 063534},
\href{http://arxiv.org/abs/0805.3781}{{\ttfamily arXiv:0805.3781 [hep-th]}}.

\bibitem{Aurenche:1993vt}
P.~Aurenche, T.~Becherrawy, and E.~Petitgirard, ``{Retarded / advanced
  correlation functions and soft photon production in the hard loop
  approximation},''
\href{http://arxiv.org/abs/hep-ph/9403320}{{\ttfamily arXiv:hep-ph/9403320
  [hep-ph]}}.

\bibitem{vanEijck:1994rw}
M.~A. van Eijck, R.~Kobes, and C.~G. van Weert, ``{Transformations of real time
  finite temperature Feynman rules},''
  \href{http://dx.doi.org/10.1103/PhysRevD.50.4097}{{\em Phys. Rev.} {\bfseries
  D50} (1994) 4097--4109},
\href{http://arxiv.org/abs/hep-ph/9406214}{{\ttfamily arXiv:hep-ph/9406214
  [hep-ph]}}.

\bibitem{Chen:2015dga}
X.~Chen, M.~H. Namjoo, and Y.~Wang, ``{On the equation-of-motion versus in-in
  approach in cosmological perturbation theory},''
  \href{http://dx.doi.org/10.1088/1475-7516/2016/01/022}{{\em JCAP} {\bfseries
  1601} no.~01, (2016) 022},
\href{http://arxiv.org/abs/1505.03955}{{\ttfamily arXiv:1505.03955
  [astro-ph.CO]}}.

\bibitem{Epelbaum:2014yja}
T.~Epelbaum, F.~Gelis, and B.~Wu, ``{Nonrenormalizability of the classical
  statistical approximation},''
  \href{http://dx.doi.org/10.1103/PhysRevD.90.065029}{{\em Phys. Rev.}
  {\bfseries D90} no.~6, (2014) 065029},
\href{http://arxiv.org/abs/1402.0115}{{\ttfamily arXiv:1402.0115 [hep-ph]}}.

\bibitem{Greene:1997fu}
P.~B. Greene, L.~Kofman, A.~D. Linde, and A.~A. Starobinsky, ``{Structure of
  resonance in preheating after inflation},''
  \href{http://dx.doi.org/10.1103/PhysRevD.56.6175}{{\em Phys. Rev.} {\bfseries
  D56} (1997) 6175--6192},
\href{http://arxiv.org/abs/hep-ph/9705347}{{\ttfamily arXiv:hep-ph/9705347
  [hep-ph]}}.

\bibitem{Dusling:2010rm}
K.~Dusling, T.~Epelbaum, F.~Gelis, and R.~Venugopalan, ``{Role of quantum
  fluctuations in a system with strong fields: Onset of hydrodynamical flow},''
  \href{http://dx.doi.org/10.1016/j.nuclphysa.2010.11.009}{{\em Nucl. Phys.}
  {\bfseries A850} (2011) 69--109},
\href{http://arxiv.org/abs/1009.4363}{{\ttfamily arXiv:1009.4363 [hep-ph]}}.

\bibitem{flagolet}
P.~Flajolet and R.~Sedgewick, ``{Analytic Combinatorics},'' {\em Cambridge
  University Press} (2009) .

\end{thebibliography}\endgroup

\providecommand{\href}[2]{#2}\begingroup\raggedright\endgroup

\end{document}